\documentclass[iop,usenatbib]{emulateapj}

\usepackage[english]{babel}
\usepackage{graphicx}
\usepackage{fleqn}
\usepackage{amssymb}
\usepackage{amsmath}
\usepackage{booktabs}
\usepackage{comment}

\usepackage[breaklinks,colorlinks,
   urlcolor=blue,citecolor=blue,linkcolor=blue]{hyperref}
\usepackage{comment}

\usepackage{color, colortbl}

\usepackage{txfonts}

\newcommand*{\mycdot}{\kern-.2em\cdot\kern-.2em}
\renewcommand{\S}{Section}
\newcommand{\F}{Fig.}
\newcommand{\codename}{SecularMultiple}

\newcommand{\msun}{\mathrm{M}_\odot}

\newcommand{\au}{\,\textsc{au}}

\newcommand{\kms}{\mathrm{km \, s^{-1}}}
\newcommand{\pgpy}{\mathrm{Gpc^{-3}\,yr^{-1}}}

\newcommand\kick{\mathrm{k}}
\newcommand{\sk}{\sigma_\kick}

\newcommand{\alhs}{\alpha_{\mathrm{s}}}
\newcommand{\alhb}{\alpha_{\mathrm{bin}}}
\newcommand{\alht}{\alpha_{\mathrm{tr}}}
\newcommand{\nsys}{N_{\mathrm{sys}}}

\allowdisplaybreaks

\begin{document}

\title{Double neutron star mergers from hierarchical triple-star systems}
\author{Adrian S. Hamers$^{1}$ and Todd A. Thompson$^{2,3,1}$}
\affil{$^{1}$Institute for Advanced Study, School of Natural Sciences, Einstein Drive, Princeton, NJ 08540, USA \\
$^{2}$Department of Astronomy, The Ohio State University, Columbus, Ohio 43210, USA \\
$^{3}$Center for Cosmology and AstroParticle Physics, Department of Physics, The Ohio State University, Columbus, Ohio 43210, USA}
\email{hamers@ias.edu}



\begin{abstract} 
The isolated binary evolution model for merging neutron stars (NSs) involves processes such as mass transfer, common-envelope evolution, and natal kicks, all of which are poorly understood. Also, the predicted NS-NS merger rates are typically lower than the rates inferred from the LIGO GW170817 event. Here, we investigate merger rates of NS and black hole (BH)-NS binaries in hierarchical triple-star systems. In such systems, the tertiary can induce Lidov-Kozai (LK) oscillations in the inner binary, accelerating its coalescence, and potentially enhancing compact object merger rates. However, since compact objects originate from massive stars, the prior evolution should also be taken into account. Natal kicks, in particular, could significantly reduce the rates by unbinding the tertiary before it can affect the inner binary through LK evolution. We carry out simulations of massive triples taking into account stellar evolution starting from the main sequence, secular and tidal evolution, and the effects of supernovae. For large NS birth kicks ($\sigma_\kick=265\,\kms$), we find that the triple NS-NS merger rate (several hundred $\pgpy$) is lower by a factor of $\sim2-3$ than the binary rate, but for no kicks ($\sigma_\kick=0\,\kms$), the triple rate (several thousand $\pgpy$) is comparable to the binary rate. Our results indicate that a significant fraction of NS-NS mergers could originate from triples if a substantial portion of the NS population is born with low kick velocities, as indicated by other work. However, uncertainties and open questions remain because of our simplifying assumption of dynamical decoupling after inner binary interaction has been triggered. 
\end{abstract}

\section{Introduction}
\label{sect:introduction}
The detection of a double neutron star (NS) merger by LIGO's O2 observing run \citep{2017PhRvL.119p1101A} has definitively shown that double neutron stars (NSs) can merge in the Universe, and the accompanying electromagnetic signals in gamma rays \citep[e.g.,][]{2017ApJ...848L..13A,2017ApJ...848L..14G,2017ApJ...848L..15S}, X-rays \citep[e.g.,][]{2017ApJ...848L..20M}, optical \citep[e.g.,][]{2017Sci...358.1556C,2017Sci...358.1574S,2017ApJ...848L..16S,2017ApJ...848L..18N}, near-infrared \citep[e.g.,][]{2017ApJ...848L..19C,2017ApJ...848L..17C,2017Natur.551...67P}, and radio wavelengths \citep[e.g.,][]{2017ApJ...848L..21A}, have revealed a wealth of information on kilonova transients \citep[e.g.,][]{1974ApJ...192L.145L,1999ApJ...525L.121F,2010MNRAS.406.2650M,2015IJMPD..2430012R,2015MNRAS.448..541J}, and short gamma-ray bursts \citep[e.g.,][]{1989Natur.340..126E,1999PhR...314..575P,2014ARA&A..52...43B}. The origin of compact object mergers is unclear, since in standard stellar evolution theory, the progenitor stars would merge before evolving to compact objects in an orbit that would cause them to merge due to gravitational wave (GW) emission within a Hubble time. Two main scenarios have been proposed for compact object mergers: isolated binary evolution \citep[e.g.,][]{1973NInfo..27...70T,1992ApJ...386..197T,1993MNRAS.260..675T,1997AstL...23..492L,2002ApJ...572..407B,2003MNRAS.342.1169V,2007PhR...442...75K,2012ApJ...759...52D,2013ApJ...779...72D,2014ApJ...789..120B,2016Natur.534..512B,2017arXiv170607053B,2017NatCo...814906S,2018MNRAS.474.2937C,2019MNRAS.482.2234G} and dynamical interactions, such as those in triple-star systems \citep[e.g.,][]{2011ApJ...741...82T,2013MNRAS.430.2262H,2017ApJ...841...77A,2017ApJ...836...39S,2017ApJ...846L..11L,2018ApJ...863...68L,2018A&A...610A..22T,2019arXiv190500427L}, globular clusters \citep[e.g.,][]{1993Natur.364..423S,2000ApJ...528L..17P,2006ApJ...637..937O,2014MNRAS.441.3703Z,2015PhRvL.115e1101R,2016PhRvD..93h4029R,2016MNRAS.463.2443K,2016MNRAS.459.3432M,2017ApJ...840L..14S,2018ApJ...855..124S,2018ApJ...853..140S,2018PhRvD..97j3014S,2018PhRvL.120o1101R}, and galactic nuclei \citep[e.g.,][]{2012ApJ...757...27A,2014ApJ...781...45A,2015ApJ...799..118P,2016MNRAS.460.3494S,2016ApJ...828...77V,2016ApJ...831..187A,2017ApJ...846..146P,2018MNRAS.477.4423A,2018ApJ...864..134R,2018ApJ...853...93R,2018ApJ...865....2H,2018ApJ...860....5G,2018ApJ...856..140H,2018arXiv181110627F,2019MNRAS.483..152A}. 

The isolated binary channel involves binary interactions such as mass transfer, common-envelope (CE) evolution, and supernovae (SNe) kicks associated with the birth of NSs or black holes (BHs). The details of these processes are highly uncertain, yet, taking into account these uncertainties, it is challenging in the isolated binary model to predict rates (e.g., \citealt{2004ApJ...601L.179K,2004ApJ...614L.137K}, see also \citealt{2010CQGra..27q3001A} for a review) that are on the same order of magnitude as derived from the LIGO observation of GW170817, $1540_{-1220}^{+3200}\,\pgpy$ \citep{2017PhRvL.119p1101A}. For example, \citet{2018MNRAS.480.2011G} find rates of up to $1\times10^3 \,\pgpy$, but only assuming low SNe kicks, and a high CE efficiency.

Triple systems have been invoked in a large variety of contexts to explain an enhanced rate of interaction in binary systems (see, e.g., \citealt{2016ARA&A..54..441N} for a review). This occurs through oscillations of the inner binary eccentricity as a result of the torque of the tertiary, known as Lidov-Kozai (LK) oscillations \citep{1962P&SS....9..719L,1962AJ.....67..591K}. In particular, in the context of compact objects, it has been suggested that LK-driven eccentricity excitation can accelerate the mergers of compact objects \citep[e.g.,][]{2002ApJ...578..775B,2002ApJ...576..894M,2003ApJ...598..419W,2011ApJ...741...82T,2017ApJ...841...77A,2018ApJ...856..140H,2018PASA...35...17B,2019ApJ...878...58S,2019MNRAS.486.4443F}. However, since compact objects originate from massive stars\footnote{Excluding the possibility of primordial compact objects.}, the prior stellar evolution should also be taken into account, similarly to studies of binary evolution \citep[e.g.,][]{2013MNRAS.430.2262H,2016MNRAS.460.3494S,2016ApJ...822L..24N,2018A&A...610A..22T,2018MNRAS.478..620H,2019ApJ...878...58S}. Natal kicks, in particular, could significantly reduce the rates by unbinding the tertiary before it can affect the inner binary through LK evolution \citep[e.g.,][]{2019MNRAS.484.1506L,2019arXiv190412881H}. 

In this paper, we investigate the interplay between these processes in massive triple-star systems, taking into account stellar evolution starting from the main sequence (MS), secular and tidal evolution, and the effects SNe. In contrast to previous studies which focused on wide inner binaries that do not interact in the absence of a tertiary star \citep[e.g.,][]{2013MNRAS.430.2262H,2017ApJ...841...77A,2019arXiv190412881H}, we here consider triples with no restrictions on the inner binary orbital separation. In particular, this implies that in the `binary case' of our triples, i.e., when the effect of the tertiary on the inner binary is ignored, the system can in fact interact, and possibly produce a double NS merger via the standard binary evolution channel. 

This approach requires to model binary processes such as mass transfer and CE evolution, as well as the effects associated with the gravitational perturbations from the tertiary (i.e., LK oscillations). Combining these processes is challenging and complicated (see, e.g., \citealt{2019ApJ...872..119H} for an exploratory study for the case of mass transfer taking into account orbital effects due to mass transfer and LK evolution). Here, we take a simpler approach in which we model the evolution of the tertiary initially using a secular code taking into account dynamical, stellar and tidal evolution, but not mass transfer nor CE evolution. We track the onset of mass transfer (i.e., Roche-Lobe overflow, RLOF), and in this case, we continue the evolution of the inner binary system using a dedicated binary population synthesis code which includes all the required binary physics, but not any effects associated with the tertiary star. Here, we make the assumption that, after the onset of RLOF in the inner binary, the latter is dynamically decoupled from the tertiary. This assumption is typically justified in the case of CE evolution, in which case the inner binary shrinks significantly (see Fig. 8 of \citealt{2013MNRAS.430.2262H}). However, in the case of mass transfer, the inner orbit can also expand (if the donor has become less massive than the accretor). We here ignore this complication, but instead take the simpler, decoupled approach. 

Several types of compact object mergers have been studied in triple systems, including WD-WD mergers \citep{2011ApJ...741...82T,2012arXiv1211.4584K,2013MNRAS.430.2262H,2018A&A...610A..22T,2019ApJ...878...58S}, BH-BH mergers \citep{2017ApJ...836...39S,2017ApJ...841...77A,2017ApJ...846L..11L,2018ApJ...863...68L,2018ApJ...856..140H,2019arXiv190500427L}, and BH-NS mergers \citep{2019ApJ...878...58S,2019MNRAS.486.4443F}. To our knowledge, NS-NS mergers in triples with stellar-mass tertiaries (taking into account stellar evolution and dynamical evolution; see \citealt{2019ApJ...878...58S} for a comparable study, but with supermassive BH tertiaries) have not been studied. Our focus is therefore on mergers of double NSs. 

The structure of this paper is as follows. We describe our methodology in \S~\ref{sect:meth}, and the initial conditions of our simulations in \S~\ref{sect:IC}. We present our results, most notably the merger rates, in \S~\ref{sect:results}. We discuss our findings in \S~\ref{sect:discussion}, and conclude in \S~\ref{sect:conclusions}.

\section{Numerical method}
\label{sect:meth}
We use a hybrid method in which we use \textsc{\codename} \citep{2016MNRAS.459.2827H,2018MNRAS.476.4139H} to model the secular dynamical, tidal, and stellar evolution of a binary or hierarchical triple star system, and the binary stellar evolution code \textsc{BSE} \citep{2000MNRAS.315..543H,2002MNRAS.329..897H} to model the evolution of systems in which we consider the tertiary to be unimportant. The latter case, which we will refer to as `isolated' binary evolution, includes interacting systems that undergo mass transfer in the inner binary, and systems in which the tertiary star becomes unbound from the inner binary due to a SNe event but with the inner binary remaining bound, and possibly merging at a later time. Both codes, \textsc{\codename} and \textsc{BSE}, are implemented within the \textsc{AMUSE} framework \citep{2013CoPhC.183..456P,2013A&A...557A..84P}. 

\subsection{\textsc{\codename}}
\label{sect:meth:sm}
In \textsc{\codename}, we model the evolution of a binary or triple system, starting from MS stars, and taking into account stellar evolution, tidal evolution, and secular dynamical evolution (in the case of triples, and up to and including fifth order in the expansion of the separation ratio of the inner to the outer binary, see \citealt{2016MNRAS.459.2827H,2018MNRAS.476.4139H}\footnote{The reference frame in \textsc{\codename} is an arbitrary frame, rather than the invariable plane.}). The modeling in \textsc{\codename} is similar to that in previous works in which we coupled secular dynamical evolution to stellar and tidal evolution \citep{2016MNRAS.462L..84H,2018MNRAS.478..620H,2019arXiv190412881H}.

Stellar evolution is taken into account by using \textsc{SSE} \citep{2000MNRAS.315..543H} (as implemented in \textsc{AMUSE}), which is based on analytical fits to detailed stellar evolution models, and which uses the same stellar tracks as (i.e., is consistent with) \textsc{BSE}. We set the metallicity to Solar, $Z=0.02$. Quantities that are used from \textsc{SSE} include the stellar type and (convective envelope) mass and radius as a function of age. These quantities are used to take into account mass loss from the system, assumed to occur either adiabatically, or because of an impulsive change due to SNe. In the former case, we assume isotropic and adiabatic mass loss, i.e., $a_i M_i$ and $e_i$ are constant \citep{1956AJ.....61...49H,1963ApJ...138..471H} for an orbit $i$ in the system ($i=1$ and $i=2$, for the inner and outer orbits, respectively, and if applicable). Here, $M_1=m_1+m_2$ for the inner orbit, and $M_2=M_1+m_3$ for the outer orbit. 

In the case of SNe, we use the routines incorporated into \textsc{\codename} and described in \citet{2018MNRAS.476.4139H} to compute the effects on the inner and outer orbits of the (assumed to be instantaneous) mass loss and (possible) kick to the newly-formed NS or BH, assuming a random orbital phase of both orbits at the moment of SNe. We assume that the kick distribution is a Maxwellian distribution, i.e., the probability density function of the kick speed $V_\kick$ is given by
\begin{align}
\label{eq:v_kick}
\frac{\mathrm{d}N}{\mathrm{d} V_\kick} = \sqrt{\frac{2}{\pi}} \frac{V_\kick^2}{\sk^3} \exp \left ( -\frac{V_\kick^2}{2\sk^2} \right ).
\end{align}
Here, $\sk$ characterizes the typical kick speed, and is assumed to be given (see \S\,\ref{sect:IC} below). For BHs, we use the prescription of \citet[Section 4.1]{2012ApJ...749...91F} to rescale the sampled speed $V_\kick$ in equation~(\ref{eq:v_kick}) according to $V_{\kick,\,\mathrm{BH}} = V_\kick (1-f_{\mathrm{fb}})$, where the fallback factor $f_{\mathrm{fb}}$ is given by
\begin{align}
\label{eq:v_kick_bh}
f_{\mathrm{fb}} = \left \{ \begin{array}{cc}
0, & M_{\mathrm{CO}} < 5 \, \msun; \\
0.378 M_{\mathrm{CO}} - 1.889, & 5.0\,\msun \leq M_{\mathrm{CO}} < 7.6\,\msun; \\
1, & M_{\mathrm{CO}} \geq 7.6\,\msun,
\end{array} \right.
\end{align}
where $M_{\mathrm{CO}}$ is the mass of the carbon-oxygen core of the proto-BH, which is extracted from \textsc{SSE}.

Tidal evolution is taken into account with the assumption of the equilibrium tide model \citep{1981A&A....99..126H,1998ApJ...499..853E}. Specifically, we use equations (81) and (82) of \citet{1998ApJ...499..853E}, with the non-dissipative terms $X$, $Y$ and $Z$ given explicitly by equation (10)-(12) of \citet{2001ApJ...562.1012E}, and the dissipative terms $V$ given explicitly by equations (A7)-(A11) of \citet{2009MNRAS.395.2268B}. We apply these terms to both the inner and outer orbits (if applicable), in the latter case treating the inner binary as a point mass. The stellar spins are included in the tidal calculations; the initial spin-orbit angles are assumed to be zero (i.e., zero obliquities). The tidal dissipation efficiency is computed as a function of the stellar parameters as part of the set of ordinary differential equations using the prescription of \citet{2002MNRAS.329..897H}. For all stars, we assume a fixed apsidal motion constant of $k_{\mathrm{AM},i}=0.014$, and a gyration radius of 0.08 (these parameters are not provided by \textsc{SSE} in \textsc{AMUSE}). 

Many uncertainties still remain in tidal evolution (see, e.g., \citealt{2014ARA&A..52..171O} for a review). Especially for high eccentricities, which could be excited by LK oscillations, the equilibrium tide model could break down. A more sophisticated treatment of tides is beyond the scope of this work.

During the integration of the system in \textsc{\codename}, we check for a number of stopping conditions. These are listed and explained below.
\begin{enumerate}
\item For both binaries and triples, we check for RLOF in the inner binary using analytic fits \citep[eqs. 47-52] {2007ApJ...660.1624S}. Specifically, we check for RLOF at periapsis as a function of the current radii, spins, and inner binary semimajor axis and eccentricity. If RLOF occurs, we stop the simulation within \textsc{\codename}, and continue the evolution of the inner binary using \textsc{BSE}; see \S~\ref{sect:meth:bse} below for further details. We also check for RLOF by the tertiary star on the inner binary, treating the binary as a point mass and using the same fits of \citet{2007ApJ...660.1624S}, although this scenario occurs much less commonly compared to RLOF in the inner binary (see Tables~\ref{table:fractions_q2} and \ref{table:fractions_q2p}). 
\item For both binaries and triples, after each SNe event, we check if the system is still bound (i.e,. if $a_i>0$ for $i=1$ and $i=2$ if applicable). If not, we continue the evolution of the inner binary with \textsc{BSE} if the outer orbit became unbound, but the inner binary is still bound (and could interact and merge at a later time); see \S~\ref{sect:meth:bse} below.
\item For triples, we check for dynamical stability of the system using the criterion of \citet{2001MNRAS.321..398M}, and stop the simulation in \textsc{\codename} if the system is unstable according to this criterion. We do not consider the subsequent evolution of the system. 
\item For triples, we check if at any point in the evolution if the secular approximation made in the equations of motion breaks down. Such a breakdown can occur if the timescale for the inner binary angular momentum to change significantly becomes comparable to the inner or outer orbital periods (e.g., \citealt{2004AJ....128.2518C,2005MNRAS.358.1361I,2012arXiv1211.4584K,2012ApJ...757...27A,2013PhRvL.111f1106S,2014ApJ...781...45A,2014MNRAS.439.1079A,2014MNRAS.438..573B,2016MNRAS.458.3060L,2018MNRAS.481.4602L,2018MNRAS.481.4907G}). To evaluate whether or not this has occurred, we use the criterion of \citet{2014ApJ...781...45A}, i.e.,
\begin{align}
\sqrt{1-e_1} < 5 \pi \frac{m_3}{m_1+m_2} \left [ \frac{a_1}{a_2 \left(1-e_2 \right )} \right ]^3.
\end{align}
Although we check for this regime (also known as the semisecular regime), it occurs very rarely in our simulations since extremely high eccentricities are required, and in most cases this implies that the stars in the inner binary would undergo mass transfer well before. 
\item The inner binary components collide directly, i.e., $a_1(1-e_1)<R_1+R_2$, where $R_1$ and $R_2$ are the primary and secondary radius, respectively. This typically can only occur for compact objects, since the larger sizes of pre-compact objects generally imply that, when the eccentricity is high, usually RLOF occurs or possibly the semisecular regime is triggered. 
\item The system exceeds an age of 10 Gyr. 
\end{enumerate}

In contrast to previous work related to white dwarfs \citep{2018MNRAS.478..620H,2019arXiv190412881H}, we here do not include the effects of flybys on the binary or triple system. We find that most NS-NS mergers occur relatively early, i.e., within $\sim 100\,\mathrm{Myr}$ (see \S~\ref{sect:results} below). For systems in the field, flybys are typically unimportant on such time-scales. Mergers occurring as a result of perturbations from passing stars are beyond the scope of this paper (see, e.g., \citealt{2014ApJ...782...60K,2019arXiv190201864M}).

\subsection{\textsc{BSE}}
\label{sect:meth:bse}
As described in \S~\ref{sect:meth}, we consider two situations in the simulations with \textsc{\codename} in which we assume for triples that the inner binary can subsequently be decoupled from the tertiary star. We then continue the evolution with the binary population synthesis code \textsc{BSE}. The latter code includes prescriptions for binary-star evolution, most notably mass transfer and CE evolution, which are both not modeled in \textsc{\codename}. In \textsc{BSE}, we assume a CE binding energy factor of $\lambda=0.5$, and a CE efficiency of $\alpha=1$. Furthermore, in \textsc{BSE} we assume the same NS kick speed distribution as in \textsc{\codename} (see equation~\ref{eq:v_kick}). Due to limitations of \textsc{BSE}, we do not include the fallback correction to kicks applied to BHs, as described by equation~(\ref{eq:v_kick_bh}). Instead, in \textsc{BSE}, we assume zero kick speeds for BHs. This is strictly inconsistent with the prior simulations in \textsc{\codename}, but our main focus in this work is on merging NSs, so we do not expect this inconsistency to strongly affect our conclusions. 

The two situations in which we hand off the evolution in \textsc{\codename} to \textsc{BSE} correspond to stopping conditions (1) and (2) described in \S~\ref{sect:meth:sm}. In situation (1), the inner binary undergoes RLOF, which can be triggered by several factors: expansion of the stars in the inner binary due to stellar evolution, possibly combined with tidal evolution, and/or high eccentricity due to secular evolution. Note that, contrary to previous studies, we do not restrict to wide systems. Therefore, many systems will undergo RLOF even in the `binary' case, i.e., in absence of a tertiary star. 

The crucial assumption made in situation (1) for triples is that the subsequent evolution of the inner binary is completely decoupled from the tertiary star. This is a strong assumption, and is certainly not correct for all systems. The motivation for this approach is simplicity, since we currently do not have the tools to self-consistently simulate the evolution of the triple system taking into account the secular dynamical evolution with mass transfer, in particular if the orbit is eccentric and driven to varying eccentricity due to LK oscillations with similar RLOF and LK time-scales. Preliminary work \citep{2019ApJ...872..119H} indicates that the evolution in this regime is complicated, and merits future study. The decoupling assumption likely gives rise to a significant systematic uncertainty in our results. We further discuss this issue in \S~\ref{sect:discussion:decouple}.

\section{Initial conditions}
\label{sect:IC}
Here, we describe the initial conditions assumed in our simulations. We consider several sets of initial conditions; in each set, we generate $N_{\mathrm{MC}} = 10^3$ systems through Monte Carlo sampling. Each set is characterized by the type of system -- `Triple' or `Binary', where `Binary' refers to the same system, but in absence of the tertiary star --, the dispersion $\sk$ assumed in the kick speed distribution (see equation~\ref{eq:v_kick}), and the type of assumption on the tertiary mass distribution (see \S~\ref{sect:IC:mass} below). 

\begin{table*}
\begin{tabular}{lp{6.0cm}p{8.0cm}}
\toprule
Symbol & Description & Initial value(s) and/or distribution in population synthesis \\
\midrule
$m_1$					& Mass of the primary star.																	& $8-50\,\msun$ with a Salpeter initial mass function (\citealt{1955ApJ...121..161S}, i.e., $\mathrm{d} N/\mathrm{d}m_1 \propto m_1^{-2.35}$). \\
$m_2$					& Mass of the secondary star. 																	& $m_1 q_1$, where $q_1$ has a flat distribution, and with $m_2>4\,\msun$. \\
$m_3$ 					& Mass of the tertiary star.																		& Either (1) $m_3 = q_2 (m_1+m_2)$, or (2) $m_3=q_2' m_2$, where both $q_2$ and $q_2'$ have a flat distribution, subject to $m_3>0.1\,\msun$. \\
$Z_i$					& Metallicity of star $i$.																		& $0.02$ \\
$R_i$					& Radius of star $i$.																			& From stellar evolution code. \\
$P_{\mathrm{s},i}$			& Spin period of star $i$. 																		& $10\,\mathrm{d}$ \\
$\theta_{\mathrm{s},i}$		& Obliquity (spin-orbit angle) of star $i$. 															& $0^\circ$ \\
$t_{\mathrm{V},i}$			& Viscous time-scale of star $i$. 																& Computed from the stellar properties using the prescription of \citet{2002MNRAS.329..897H}. \\
$k_{\mathrm{AM},i}$			& Apsidal motion constant of star $i$.															& 0.014 \\
$r_{\mathrm{g},i}$			& Gyration radius of star $i$.																	& 0.08 \\
$V_\kick$					& SNe kick speed.																			& Maxwellian distribution with $\sk=0$, 40, or $265\,\kms$. \\
$P_{\mathrm{orb},i}$			& Orbital period of orbit $i$ (inner orbit: $i=1$; outer orbit: $i=2$).										& Flat distribution in $\log_{10}(P_{\mathrm{orb},i})$, with $1<\log_{10}(P_{\mathrm{orb},i}/\mathrm{d})<10$, and subject to dynamical stability constraints. \\
$a_i$					& Semimajor axis of orbit $i$. 																	& Computed from $P_{\mathrm{orb},i}$ and the $m_i$ using Kepler's law. \\
$e_i$					& Eccentricity of orbit $i$. 																		& Flat distribution between 0 and 0.9. \\
$i_i$						& Inclination of orbit $i$. 																		& $0-180^\circ$ (flat distribution in $\cos i_i $) \\
$\omega_i$				& Argument of periapsis of orbit $i$. 																& $0-360^\circ$ (flat distribution in $\omega_i$) \\
$\Omega_i$				& Longitude of the ascending node of orbit $i$. 														& $0-360^\circ$ (flat distribution in $\Omega_i$) \\
\bottomrule
\end{tabular}
\caption{Description of important quantities and their initial value(s) and/or distributions assumed in the population synthesis. }
\label{table:IC}
\end{table*}

Below, we describe in more detail the assumptions made in the Monte Carlo sampling. An overview of our notation and assumptions is given in Table~\ref{table:IC}.

\subsection{Masses}
\label{sect:IC:mass}
We sample the mass of the primary star, $m_1$, from a Salpeter distribution \citep{1955ApJ...121..161S}, i.e., $\mathrm{d}N/\mathrm{d}m_1 \propto m_1^{-2.35}$. Since our interest is in NS mergers, we sample with the range $8<m_1/\msun<50$. The mass of the secondary star, $m_2$ is sampled assuming a flat distribution in $q_1 \equiv m_2/m_1$, consistent with observations of massive stars \citep[e.g.,][]{2012Sci...337..444S,2013ARA&A..51..269D,2014ApJS..213...34K}. Here, we set the lower limit on $m_2$ to be $4\,\msun$. The tertiary mass $m_3$ is sampled according to two methods: sampling from (1) a flat distribution in the mass ratio $q_2 \equiv m_3/(m_1+m_2)$, and (2) a flat distribution in the mass ratio $q_2' \equiv m_3/m_2$. In either case, we set the lower limit on $m_3$ to be $0.1\,\msun$. These two choices give different results in our simulations, since in case (1) the tertiary star can be more massive than the primary star, and therefore can evolve first. Consequently, the tertiary star can become unbound from the inner binary as it explodes in a SNe event. In contrast, in method (2), the tertiary star is always less massive than the primary star, such that the inner binary always evolves first. 

\begin{figure}
\center
\includegraphics[scale = 0.45, trim = 0mm 0mm 0mm 0mm]{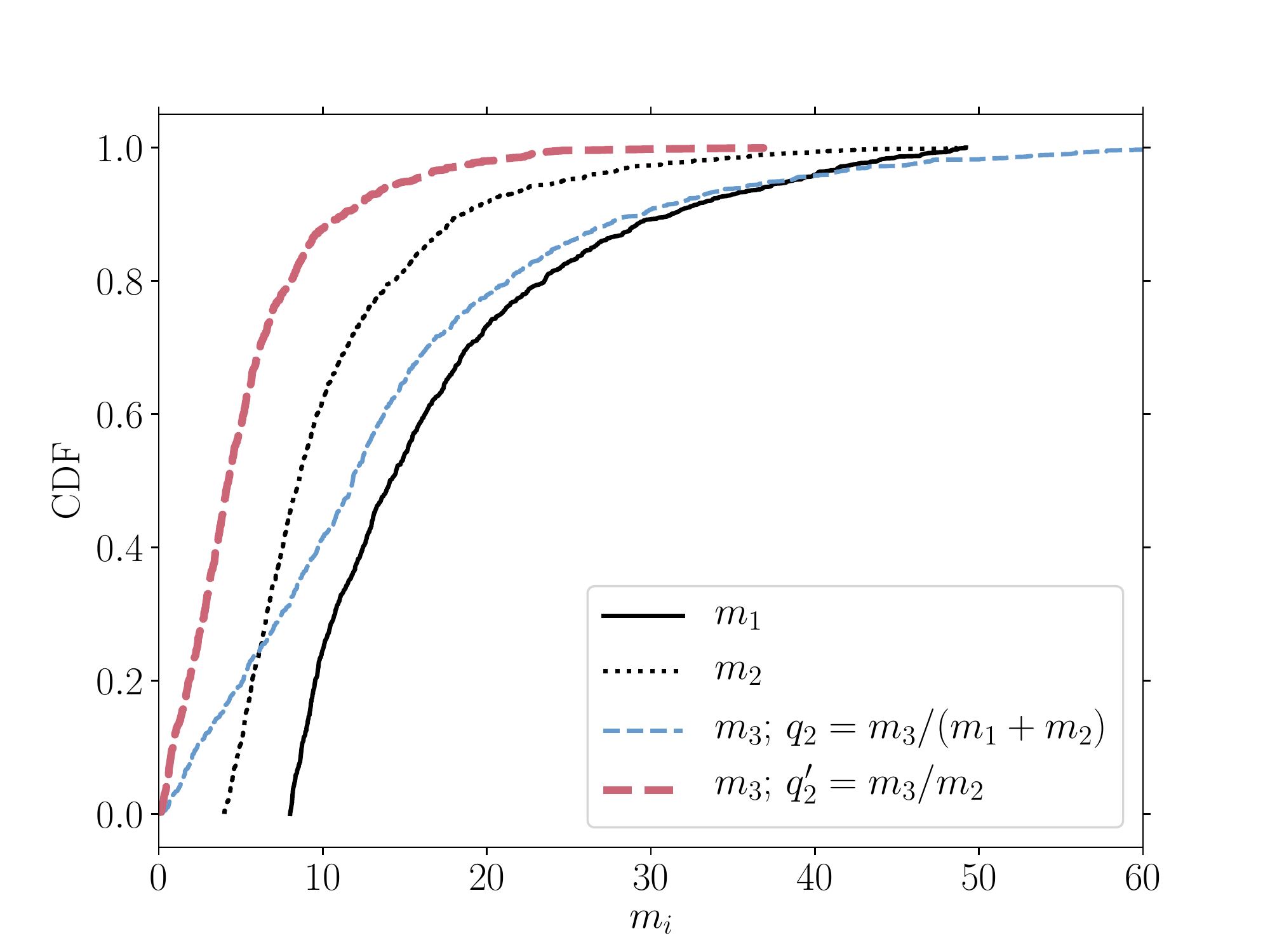}
\caption { Distributions of the initial primary, secondary, and tertiary masses in the Monte Carlo simulations. The thin (thick) dashed lines show the distributions of the tertiary mass $m_3$ assuming method 1 (2) for the tertiary mass ratio $q_2$ ($q_2'$).  }
\label{fig:IC_masses}
\end{figure}

In \F~\ref{fig:IC_masses}, we illustrate the difference between the two assumptions on the tertiary mass ratio distribution. With method (1), the median tertiary mass is $\approx 5\,\msun$, whereas with method (2) it is $\approx 12\,\msun$. 

\subsection{Orbits}
\label{sect:IC:orbits}
For both inner and outer orbits, we assume a flat distribution in the logarithm of the orbital period, i.e., flat in $\log_{10}(P_{\mathrm{orb},i})$ (also known as an \"{O}pic distribution, \citealt{1924PTarO..25f...1O}). The limits for both inner and outer orbits are set to $1<\log_{10}(P_{\mathrm{orb},i}/\mathrm{d})<10$. The eccentricities $e_i$ of both inner and outer orbits are sampled from flat distributions in $e_i$, with $0<e_j<0.9$. The orbital period and eccentricity distributions are roughly consistent with observations of massive binary stars \citep{2014ApJS..213...34K}. Given the uncertainties in the observed orbital distributions of massive triples, we do not take into account a more sophisticated initial distribution, even though the observed distribution for massive MS triples reflects the initial distribution \citep{2019MNRAS.488.2480R}. 

We reject a sampled system if it is unstable according to the criterion of \citet{2001MNRAS.321..398M}. The orientations of the orbits are taken to be random, i.e., the inclinations $i_i$ are sampled from flat distributions in $\cos i_i$, and the arguments of periapsis $\omega_i$ and longitudes of the ascending node $\Omega_i$ are sampled from flat distributions. There are indications that lower-mass triples have more aligned inner and outer orbits if the system is compact ($a_2 \lesssim 100\,\au$), whereas the wider systems are more isotropically distributed \citep{2017ApJ...844..103T}. However, it is unclear if this trend also persists for higher-mass triples (primaries with masses $\gtrsim 8 \,\msun$). For simplicity, we here restrict to randomly-oriented triples.

\subsection{Kick distributions}
\label{sect:IC:kick}
As discussed in \S~\ref{sect:meth}, we sample the kick distribution for SNe from a Maxwellian distribution \citep[e.g.,][]{1997MNRAS.291..569H}. We take the dispersion $\sk$ to be fixed for each set of Monte Carlo simulations, and adopt three values: 0, 40, and 265 $\kms$. The value $\sk=0\,\kms$ is to evaluate the importance of mass loss associated with SNe only (i.e., only the Blaauw kick, \citealt{1961BAN....15..265B,1961BAN....15..291B}). The value $\sk=265\,\kms$ is a commonly-adopted value inferred from the proper motions of pulsars \citep{2005MNRAS.360..974H}. However, NS kicks are uncertain and their magnitude is still highly debated. For example, \citet{2002ApJ...568..289A} found a two-component distribution of kick speed distributions with characteristic velocities of $90$ and $500 \,\kms$ based on the velocities of isolated radio pulsars, and \citet{2016MNRAS.456.4089B} also found evidence for a bimodal distribution based on observed binary NSs, with a low-kick population with $V_\kick<30\,\kms$, and a high-kick population with kicks up to $400\,\kms$. Rather than adopting a bimodal distribution, we here choose to carry out another set with $\sk=40\,\kms$, to evaluate the importance of low but non-zero kicks.

\section{Results}
\label{sect:results}

\begin{table*}
\begin{tabular}{lcccccc}
\toprule
& \multicolumn{6}{c}{Fraction of all systems} \\
& \multicolumn{3}{c}{Triple} &\multicolumn{3}{c}{Binary} \\
\midrule
& \multicolumn{3}{c}{$\sk/\kms$} &\multicolumn{3}{c}{$\sk/\kms$} \\
& 0 & 40 & 265 & 0 & 40 & 265\\
\midrule
No Interaction & $0.024 \pm 0.005$ & $0.004 \pm 0.002$ & $0.004 \pm 0.002$ & $0.253 \pm 0.016$ & $0.011 \pm 0.003$ & $0.008 \pm 0.003$ \\
\midrule
RLOF $\star$1 & $0.607 \pm 0.025$ & $0.445 \pm 0.021$ & $0.442 \pm 0.021$ & $0.612 \pm 0.025$ & $0.612 \pm 0.025$ & $0.612 \pm 0.025$ \\
\quad MS & $0.068 \pm 0.008$ & $0.063 \pm 0.008$ & $0.061 \pm 0.008$ & $0.056 \pm 0.007$ & $0.056 \pm 0.007$ & $0.056 \pm 0.007$ \\
\quad G & $0.444 \pm 0.021$ & $0.314 \pm 0.018$ & $0.313 \pm 0.018$ & $0.460 \pm 0.021$ & $0.460 \pm 0.021$ & $0.460 \pm 0.021$ \\
\quad CHeB & $0.095 \pm 0.010$ & $0.068 \pm 0.008$ & $0.068 \pm 0.008$ & $0.096 \pm 0.010$ & $0.096 \pm 0.010$ & $0.096 \pm 0.010$ \\
\midrule
RLOF $\star$2 & $0.024 \pm 0.005$ & $0.021 \pm 0.005$ & $0.021 \pm 0.005$ & $0.021 \pm 0.005$ & $0.009 \pm 0.003$ & $0.002 \pm 0.001$ \\
\quad MS & $0.021 \pm 0.005$ & $0.021 \pm 0.005$ & $0.021 \pm 0.005$ & $0.000 \pm 0.000$ & $0.000 \pm 0.000$ & $0.000 \pm 0.000$ \\
\quad G & $0.002 \pm 0.001$ & $0.000 \pm 0.000$ & $0.000 \pm 0.000$ & $0.021 \pm 0.005$ & $0.009 \pm 0.003$ & $0.002 \pm 0.001$ \\
\quad CHeB & $0.001 \pm 0.001$ & $0.000 \pm 0.000$ & $0.000 \pm 0.000$ & $0.000 \pm 0.000$ & $0.000 \pm 0.000$ & $0.000 \pm 0.000$ \\
\midrule
RLOF $\star$3 & $0.009 \pm 0.003$ & $0.009 \pm 0.003$ & $0.008 \pm 0.003$ & --- & --- & ---  \\
\quad MS & $0.000 \pm 0.000$ & $0.000 \pm 0.000$ & $0.000 \pm 0.000$ & --- & --- & ---  \\
\quad G & $0.004 \pm 0.002$ & $0.004 \pm 0.002$ & $0.004 \pm 0.002$ & --- & --- & ---  \\
\quad CHeB & $0.005 \pm 0.002$ & $0.005 \pm 0.002$ & $0.004 \pm 0.002$ & --- & --- & ---  \\
\midrule
Dynamical inst. & $0.028 \pm 0.005$ & $0.003 \pm 0.002$ & $0.004 \pm 0.002$ & --- & --- & ---  \\
\quad MS+MS & $0.002 \pm 0.001$ & $0.002 \pm 0.001$ & $0.003 \pm 0.002$ & --- & --- & ---  \\
\quad G+MS & $0.000 \pm 0.000$ & $0.000 \pm 0.000$ & $0.000 \pm 0.000$ & --- & --- & ---  \\
\quad CHeB+MS & $0.000 \pm 0.000$ & $0.000 \pm 0.000$ & $0.000 \pm 0.000$ & --- & --- & ---  \\
\quad NS+MS & $0.012 \pm 0.003$ & $0.000 \pm 0.000$ & $0.000 \pm 0.000$ & --- & --- & ---  \\
\quad NS+G & $0.003 \pm 0.002$ & $0.000 \pm 0.000$ & $0.000 \pm 0.000$ & --- & --- & ---  \\
\quad NS+CHeB & $0.003 \pm 0.002$ & $0.000 \pm 0.000$ & $0.000 \pm 0.000$ & --- & --- & ---  \\
\quad NS+NS & $0.000 \pm 0.000$ & $0.000 \pm 0.000$ & $0.000 \pm 0.000$ & --- & --- & ---  \\
\quad BH+MS & $0.004 \pm 0.002$ & $0.000 \pm 0.000$ & $0.000 \pm 0.000$ & --- & --- & ---  \\
\quad BH+G & $0.001 \pm 0.001$ & $0.000 \pm 0.000$ & $0.000 \pm 0.000$ & --- & --- & ---  \\
\quad BH+CHeB & $0.001 \pm 0.001$ & $0.001 \pm 0.001$ & $0.001 \pm 0.001$ & --- & --- & ---  \\
\quad BH+NS & $0.002 \pm 0.001$ & $0.000 \pm 0.000$ & $0.000 \pm 0.000$ & --- & --- & ---  \\
\quad BH+BH & $0.000 \pm 0.000$ & $0.000 \pm 0.000$ & $0.000 \pm 0.000$ & --- & --- & ---  \\
\midrule
Semisecular & $0.004 \pm 0.002$ & $0.004 \pm 0.002$ & $0.004 \pm 0.002$ & --- & --- & ---  \\
\quad MS+MS & $0.004 \pm 0.002$ & $0.004 \pm 0.002$ & $0.004 \pm 0.002$ & --- & --- & ---  \\
\midrule
Secular collision & $0.000 \pm 0.000$ & $0.000 \pm 0.000$ & $0.000 \pm 0.000$ & $0.000 \pm 0.000$ & $0.000 \pm 0.000$ & $0.000 \pm 0.000$ \\
\quad MS+MS & $0.000 \pm 0.000$ & $0.000 \pm 0.000$ & $0.000 \pm 0.000$ & $0.000 \pm 0.000$ & $0.000 \pm 0.000$ & $0.000 \pm 0.000$ \\
\midrule
Unbound (SNe) & $0.304 \pm 0.017$ & $0.514 \pm 0.023$ & $0.517 \pm 0.023$ & $0.114 \pm 0.011$ & $0.368 \pm 0.019$ & $0.378 \pm 0.019$ \\
\quad MS+MS & $0.000 \pm 0.000$ & $0.262 \pm 0.016$ & $0.265 \pm 0.016$ & $0.000 \pm 0.000$ & $0.000 \pm 0.000$ & $0.000 \pm 0.000$ \\
\quad G+MS & $0.000 \pm 0.000$ & $0.004 \pm 0.002$ & $0.004 \pm 0.002$ & $0.000 \pm 0.000$ & $0.000 \pm 0.000$ & $0.000 \pm 0.000$ \\
\quad CHeB+MS & $0.000 \pm 0.000$ & $0.017 \pm 0.004$ & $0.017 \pm 0.004$ & $0.000 \pm 0.000$ & $0.000 \pm 0.000$ & $0.000 \pm 0.000$ \\
\quad NS+MS & $0.219 \pm 0.015$ & $0.136 \pm 0.012$ & $0.136 \pm 0.012$ & $0.024 \pm 0.005$ & $0.263 \pm 0.016$ & $0.269 \pm 0.016$ \\
\quad NS+G & $0.000 \pm 0.000$ & $0.001 \pm 0.001$ & $0.001 \pm 0.001$ & $0.000 \pm 0.000$ & $0.001 \pm 0.001$ & $0.001 \pm 0.001$ \\
\quad NS+CHeB & $0.022 \pm 0.005$ & $0.020 \pm 0.004$ & $0.020 \pm 0.004$ & $0.000 \pm 0.000$ & $0.029 \pm 0.005$ & $0.029 \pm 0.005$ \\
\quad NS+NS & $0.007 \pm 0.003$ & $0.000 \pm 0.000$ & $0.000 \pm 0.000$ & $0.084 \pm 0.009$ & $0.001 \pm 0.001$ & $0.000 \pm 0.000$ \\
\quad BH+MS & $0.019 \pm 0.004$ & $0.056 \pm 0.007$ & $0.056 \pm 0.007$ & $0.000 \pm 0.000$ & $0.046 \pm 0.007$ & $0.054 \pm 0.007$ \\
\quad BH+G & $0.000 \pm 0.000$ & $0.000 \pm 0.000$ & $0.000 \pm 0.000$ & $0.000 \pm 0.000$ & $0.000 \pm 0.000$ & $0.000 \pm 0.000$ \\
\quad BH+CHeB & $0.001 \pm 0.001$ & $0.005 \pm 0.002$ & $0.005 \pm 0.002$ & $0.000 \pm 0.000$ & $0.004 \pm 0.002$ & $0.004 \pm 0.002$ \\
\quad BH+NS & $0.028 \pm 0.005$ & $0.005 \pm 0.002$ & $0.005 \pm 0.002$ & $0.006 \pm 0.002$ & $0.014 \pm 0.004$ & $0.010 \pm 0.003$ \\
\quad BH+BH & $0.008 \pm 0.003$ & $0.008 \pm 0.003$ & $0.008 \pm 0.003$ & $0.000 \pm 0.000$ & $0.010 \pm 0.003$ & $0.011 \pm 0.003$ \\
\midrule
\quad Inner bound & $0.286 \pm 0.017$ & $0.304 \pm 0.017$ & $0.295 \pm 0.017$ & --- & --- & ---  \\
\bottomrule
\end{tabular}
\caption{ Outcome fractions of the simulations with the tertiary mass assuming a flat distribution in $q_2\equiv m_3/(m_1+m_2)$. The first three data columns correspond to triple systems, whereas the last three data columns correspond to the binary case, i.e., taking the same triple systems but in the absence of the tertiary star. For each case, we show results for the three different values of the kick velocity dispersion, $\sk$. Fractions are based on $N_{\mathrm{MC}}=10^3$ simulations, and quoted errors are based on Poisson statistics. Outcomes that do not apply (e.g., dynamically unstable systems in the binary case) are marked with `---'. Some of the stellar evolutionary states are G -- giant star (including red giants and asymptotic giants), and core helium burning star (CHeB). Refer to the text in \S~\ref{sect:results:fractions:main} for a description of the different channels. }
\label{table:fractions_q2}
\end{table*}

\begin{table*}
\begin{tabular}{lcccccc}
\toprule
& \multicolumn{6}{c}{Fraction of all systems} \\
& \multicolumn{3}{c}{Triple} &\multicolumn{3}{c}{Binary} \\
\midrule
& \multicolumn{3}{c}{$\sk/\kms$} &\multicolumn{3}{c}{$\sk/\kms$} \\
& 0 & 40 & 265 & 0 & 40 & 265\\
\midrule
No Interaction & $0.026 \pm 0.005$ & $0.006 \pm 0.002$ & $0.006 \pm 0.002$ & $0.221 \pm 0.015$ & $0.011 \pm 0.003$ & $0.008 \pm 0.003$ \\
\midrule
RLOF $\star$1 & $0.618 \pm 0.025$ & $0.622 \pm 0.025$ & $0.621 \pm 0.025$ & $0.595 \pm 0.024$ & $0.612 \pm 0.025$ & $0.612 \pm 0.025$ \\
\quad MS & $0.060 \pm 0.008$ & $0.060 \pm 0.008$ & $0.061 \pm 0.008$ & $0.056 \pm 0.007$ & $0.056 \pm 0.007$ & $0.056 \pm 0.007$ \\
\quad G & $0.457 \pm 0.021$ & $0.462 \pm 0.021$ & $0.460 \pm 0.021$ & $0.448 \pm 0.021$ & $0.460 \pm 0.021$ & $0.460 \pm 0.021$ \\
\quad CHeB & $0.101 \pm 0.010$ & $0.100 \pm 0.010$ & $0.100 \pm 0.010$ & $0.091 \pm 0.010$ & $0.096 \pm 0.010$ & $0.096 \pm 0.010$ \\
\midrule
RLOF $\star$2 & $0.017 \pm 0.004$ & $0.016 \pm 0.004$ & $0.016 \pm 0.004$ & $0.018 \pm 0.004$ & $0.004 \pm 0.002$ & $0.002 \pm 0.001$ \\
\quad MS & $0.014 \pm 0.004$ & $0.014 \pm 0.004$ & $0.014 \pm 0.004$ & $0.000 \pm 0.000$ & $0.001 \pm 0.001$ & $0.000 \pm 0.000$ \\
\quad G & $0.002 \pm 0.001$ & $0.002 \pm 0.001$ & $0.002 \pm 0.001$ & $0.018 \pm 0.004$ & $0.003 \pm 0.002$ & $0.002 \pm 0.001$ \\
\quad CHeB & $0.001 \pm 0.001$ & $0.000 \pm 0.000$ & $0.000 \pm 0.000$ & $0.000 \pm 0.000$ & $0.000 \pm 0.000$ & $0.000 \pm 0.000$ \\
\midrule
RLOF $\star$3 & $0.000 \pm 0.000$ & $0.000 \pm 0.000$ & $0.000 \pm 0.000$ & --- & --- & ---  \\
\quad MS & $0.000 \pm 0.000$ & $0.000 \pm 0.000$ & $0.000 \pm 0.000$ & --- & --- & ---  \\
\quad G & $0.000 \pm 0.000$ & $0.000 \pm 0.000$ & $0.000 \pm 0.000$ & --- & --- & ---  \\
\quad CHeB & $0.000 \pm 0.000$ & $0.000 \pm 0.000$ & $0.000 \pm 0.000$ & --- & --- & ---  \\
\midrule
Dynamical inst. & $0.035 \pm 0.006$ & $0.006 \pm 0.002$ & $0.005 \pm 0.002$ & --- & --- & ---  \\
\quad MS+MS & $0.004 \pm 0.002$ & $0.004 \pm 0.002$ & $0.004 \pm 0.002$ & --- & --- & ---  \\
\quad G+MS & $0.000 \pm 0.000$ & $0.000 \pm 0.000$ & $0.000 \pm 0.000$ & --- & --- & ---  \\
\quad CHeB+MS & $0.000 \pm 0.000$ & $0.000 \pm 0.000$ & $0.000 \pm 0.000$ & --- & --- & ---  \\
\quad NS+MS & $0.013 \pm 0.004$ & $0.000 \pm 0.000$ & $0.000 \pm 0.000$ & --- & --- & ---  \\
\quad NS+G & $0.005 \pm 0.002$ & $0.000 \pm 0.000$ & $0.000 \pm 0.000$ & --- & --- & ---  \\
\quad NS+CHeB & $0.002 \pm 0.001$ & $0.000 \pm 0.000$ & $0.000 \pm 0.000$ & --- & --- & ---  \\
\quad NS+NS & $0.001 \pm 0.001$ & $0.000 \pm 0.000$ & $0.000 \pm 0.000$ & --- & --- & ---  \\
\quad BH+MS & $0.005 \pm 0.002$ & $0.001 \pm 0.001$ & $0.000 \pm 0.000$ & --- & --- & ---  \\
\quad BH+G & $0.001 \pm 0.001$ & $0.000 \pm 0.000$ & $0.000 \pm 0.000$ & --- & --- & ---  \\
\quad BH+CHeB & $0.002 \pm 0.001$ & $0.001 \pm 0.001$ & $0.001 \pm 0.001$ & --- & --- & ---  \\
\quad BH+NS & $0.002 \pm 0.001$ & $0.000 \pm 0.000$ & $0.000 \pm 0.000$ & --- & --- & ---  \\
\quad BH+BH & $0.000 \pm 0.000$ & $0.000 \pm 0.000$ & $0.000 \pm 0.000$ & --- & --- & ---  \\
\midrule
Semisecular & $0.003 \pm 0.002$ & $0.002 \pm 0.001$ & $0.002 \pm 0.001$ & --- & --- & ---  \\
\quad MS+MS & $0.002 \pm 0.001$ & $0.002 \pm 0.001$ & $0.002 \pm 0.001$ & --- & --- & ---  \\
\midrule
Secular collision & $0.000 \pm 0.000$ & $0.000 \pm 0.000$ & $0.000 \pm 0.000$ & $0.000 \pm 0.000$ & $0.000 \pm 0.000$ & $0.000 \pm 0.000$ \\
\quad MS+MS & $0.000 \pm 0.000$ & $0.000 \pm 0.000$ & $0.000 \pm 0.000$ & $0.000 \pm 0.000$ & $0.000 \pm 0.000$ & $0.000 \pm 0.000$ \\
\midrule
Unbound (SNe) & $0.301 \pm 0.017$ & $0.348 \pm 0.019$ & $0.350 \pm 0.019$ & $0.160 \pm 0.013$ & $0.373 \pm 0.019$ & $0.378 \pm 0.019$ \\
\quad MS+MS & $0.000 \pm 0.000$ & $0.000 \pm 0.000$ & $0.000 \pm 0.000$ & $0.000 \pm 0.000$ & $0.000 \pm 0.000$ & $0.000 \pm 0.000$ \\
\quad G+MS & $0.000 \pm 0.000$ & $0.000 \pm 0.000$ & $0.000 \pm 0.000$ & $0.000 \pm 0.000$ & $0.000 \pm 0.000$ & $0.000 \pm 0.000$ \\
\quad CHeB+MS & $0.000 \pm 0.000$ & $0.000 \pm 0.000$ & $0.000 \pm 0.000$ & $0.000 \pm 0.000$ & $0.000 \pm 0.000$ & $0.000 \pm 0.000$ \\
\quad NS+MS & $0.218 \pm 0.015$ & $0.245 \pm 0.016$ & $0.245 \pm 0.016$ & $0.048 \pm 0.007$ & $0.266 \pm 0.016$ & $0.268 \pm 0.016$ \\
\quad NS+G & $0.000 \pm 0.000$ & $0.001 \pm 0.001$ & $0.001 \pm 0.001$ & $0.000 \pm 0.000$ & $0.001 \pm 0.001$ & $0.001 \pm 0.001$ \\
\quad NS+CHeB & $0.024 \pm 0.005$ & $0.028 \pm 0.005$ & $0.028 \pm 0.005$ & $0.000 \pm 0.000$ & $0.028 \pm 0.005$ & $0.029 \pm 0.005$ \\
\quad NS+NS & $0.009 \pm 0.003$ & $0.000 \pm 0.000$ & $0.000 \pm 0.000$ & $0.099 \pm 0.010$ & $0.001 \pm 0.001$ & $0.000 \pm 0.000$ \\
\quad BH+MS & $0.020 \pm 0.004$ & $0.050 \pm 0.007$ & $0.053 \pm 0.007$ & $0.003 \pm 0.002$ & $0.048 \pm 0.007$ & $0.055 \pm 0.007$ \\
\quad BH+G & $0.000 \pm 0.000$ & $0.000 \pm 0.000$ & $0.000 \pm 0.000$ & $0.000 \pm 0.000$ & $0.000 \pm 0.000$ & $0.000 \pm 0.000$ \\
\quad BH+CHeB & $0.000 \pm 0.000$ & $0.004 \pm 0.002$ & $0.004 \pm 0.002$ & $0.000 \pm 0.000$ & $0.004 \pm 0.002$ & $0.004 \pm 0.002$ \\
\quad BH+NS & $0.026 \pm 0.005$ & $0.008 \pm 0.003$ & $0.008 \pm 0.003$ & $0.009 \pm 0.003$ & $0.012 \pm 0.003$ & $0.010 \pm 0.003$ \\
\quad BH+BH & $0.004 \pm 0.002$ & $0.012 \pm 0.003$ & $0.011 \pm 0.003$ & $0.001 \pm 0.001$ & $0.013 \pm 0.004$ & $0.011 \pm 0.003$ \\
\midrule
\quad Inner bound & $0.269 \pm 0.016$ & $0.012 \pm 0.001$ & $0.003 \pm 0.002$ & --- & --- & ---  \\
\bottomrule
\end{tabular}
\caption{ Similar to Table~\ref{table:fractions_q2}, here for simulations with the tertiary mass sampled assuming a flat distribution in $q_2'\equiv m_3/m_2$. }
\label{table:fractions_q2p}
\end{table*}

\subsection{Outcome fractions}
\label{sect:results:fractions}

\subsubsection{Main channels}
\label{sect:results:fractions:main}
In Tables~\ref{table:fractions_q2} and \ref{table:fractions_q2p}, we show the fractions for the main channels in our simulations assuming a flat distribution in $q_2\equiv m_3/(m_1+m_2)$ and $q_2'\equiv m_3/m_2$, respectively (henceforth, we refer to the latter two assumptions on the tertiary mass ratio as the `high mass tertiary' and `low mass tertiary' cases, respectively). These data are based on simulations with \textsc{\codename} only. The first three data columns correspond to triple systems, whereas the last three data columns correspond to the binary case, i.e., taking the same triple systems but in the absence of the tertiary star. For each case, we show results for the three different values of the kick velocity dispersion, $\sk$. 

The channels shown in these tables include `No interaction', i.e,. the triple or binary survived for 10 Gyr without triggering interaction such as mass transfer, or dynamical instability or instability as a result of SNe. In most cases, this `inert' channel is unlikely; instead, much more common are RLOF, or the unbinding of the system due to SNe. The only notable exception is for the binary case with $\sk=0\,\kms$, in which case the non-interacting fraction is $\sim 0.2$. In the equivalent triple case, the non-interacting fraction is only $\sim 0.02$, i.e., a factor 10 times smaller. This can be attributed largely due to the Blaauw kick in the inner binary which can keep the inner binary bound, but make the tertiary unbound (compare the `NS+MS' Unbound fractions between the triple and binary cases with $\sk=0\,\kms$). 

For RLOF, we distinguish between RLOF of the primary, secondary, or tertiary star (in the latter case, the inner binary is treated as a point mass in the fits of \citealt{2007ApJ...660.1624S}). RLOF is predominantly triggered by the primary star, with a fraction of $\sim 0.6$ assuming $\sk=0\,\kms$, and $\sim0.4$ assuming $\sk>0\,\kms$ for the triple systems with higher-mass tertiaries. The decrease in the RLOF fraction with increasing $\sk$ in the latter case can be ascribed to the higher fraction of unbound systems, which in turn is mostly due to kicks imparted on the tertiary star when it evolves first (see the unbound MS+MS fractions for triples in Table~\ref{table:fractions_q2}). This trend is much less pronounced in the case of lower-mass tertiaries (Table~\ref{table:fractions_q2p}) -- in this case, the tertiary, which is always as massive or less massive than the secondary star, does not evolve first, and therefore is less likely to unbind the triple system due to its SNe kick. 

In contrast to triples, for binaries, the RLOF star 1 fraction is independent of $\sk$. This can be explained by noting that for binaries, the primary star is always the most massive and evolves the fastest; whether or not star fills its Roche lobe is completely determined by the initial masses ($m_1$ and $m_2$), $a_1$, and $e_1$, and independent of $\sk$. This is no longer the case for the secondary star, e.g., RLOF of the secondary star can be triggered by a SNe event of the primary star, whose properties in turn are set by $\sk$. 

It may be surprising that the RLOF fraction is typically $\sim 0.6$, for {\it both} triple and binary cases. It might be expected that the RLOF fraction would be much higher for triple systems, since RLOF can be triggered by LK evolution. To explain this, we note that in most previous studies of the onset of RLOF in triples \citep[e.g.,][]{2013MNRAS.430.2262H,2017ApJ...841...77A,2019arXiv190412881H}, the inner binaries were assumed to be wide enough to avoid interaction in the absence of a tertiary star. In contrast, we here include tight systems as well. Therefore, a significant fraction of systems already interact in the `binary case'. In addition, as shown by the higher unbound fractions for the triple cases compared to the binary cases, orbital changes due to SNe associated with the tertiary star play an important role in triples with NSs. 

The fraction of dynamically unstable systems for triples is relatively small, with the fraction being at most $\sim 0.03$ if $\sk=0\,\kms$. Dynamical instability is typically triggered by mass loss from the inner binary, i.e., the traditional triple dynamical instability scenario \citep{2012ApJ...760...99P}. Instability can also be triggered by SNe events, but this is a rare event. The dynamical instability fraction in fact decreases with increasing $\sk$, which can be understood from the larger fraction of systems becoming unbound due to SNe before a dynamical instability can be triggered. 

As discussed in \S~\ref{sect:meth}, we also check for the semisecular regime in our simulations. This regime is triggered only very rarely, and predominantly with the inner binary consisting of two MS stars. The reason for the rarity of this channel is that very high eccentricities are required to trigger it, which usually instead lead to RLOF in eccentric orbits.  

We also check for direct collisions (indicated with `Secular collision' in the tables). These do not occur for pre-compact objects, since RLOF is expected to ensue before direct collision. However, direct collisions could occur during later stages, when compact objects have formed, and when the inner orbit is excited in eccentricity due to secular evolution \citep[e.g.,][]{2011ApJ...741...82T,2012arXiv1211.4584K}. Nevertheless, we find no such direct collisions of compact objects. This can be attributed to the high fractions of systems that undergo RLOF or become unbound. In other words, the probability that the system survives without interacting or becoming unstable due to SNe and a collision is triggered at later stages, is small. We emphasize that our simulations are limited in terms of the number of systems. We therefore cannot exclude that direct collisions would occur if $N_{\mathrm{MC}}$ were increased. However, we here focus on the largest contribution to NS-NS mergers, which we find originate from interacting or unbound systems (i.e., RLOF-induced and `unbound' mergers). The latter are discussed in further detail below. 

\subsubsection{Mergers from RLOF systems}
\label{sect:results:fractions:RLOF}

\begin{table*}
\begin{minipage}{.99\linewidth}
\begin{tabular}{lcccccc}
\toprule
& \multicolumn{6}{c}{Fraction of RLOF systems ($\star 1$ or $\star$ 2)} \\
& \multicolumn{3}{c}{Triple} &\multicolumn{3}{c}{Binary} \\
\midrule
& \multicolumn{3}{c}{$\sk/\kms$} &\multicolumn{3}{c}{$\sk/\kms$} \\
& 0 & 40 & 265 & 0 & 40 & 265 \\
\midrule
RLOF $\rightarrow$ Merger & $0.743 \pm 0.034$ & $0.835 \pm 0.042$ & $0.905 \pm 0.044$ & $0.708 \pm 0.033$ & $0.842 \pm 0.037$ & $0.914 \pm 0.039$ \\
\quad MS+MS & $0.079 \pm 0.011$ & $0.101 \pm 0.015$ & $0.102 \pm 0.015$ & $0.039 \pm 0.008$ & $0.040 \pm 0.008$ & $0.041 \pm 0.008$ \\
\quad CHeB+MS & $0.035 \pm 0.007$ & $0.030 \pm 0.008$ & $0.032 \pm 0.008$ & $0.021 \pm 0.006$ & $0.021 \pm 0.006$ & $0.020 \pm 0.006$ \\
\quad CHeB+G & $0.005 \pm 0.003$ & $0.004 \pm 0.003$ & $0.004 \pm 0.003$ & $0.003 \pm 0.002$ & $0.003 \pm 0.002$ & $0.003 \pm 0.002$ \\
\quad G+MS & $0.052 \pm 0.009$ & $0.052 \pm 0.011$ & $0.052 \pm 0.011$ & $0.063 \pm 0.010$ & $0.064 \pm 0.010$ & $0.065 \pm 0.010$ \\
\quad G+CHeB & $0.000 \pm 0.000$ & $0.000 \pm 0.000$ & $0.000 \pm 0.000$ & $0.000 \pm 0.000$ & $0.000 \pm 0.000$ & $0.000 \pm 0.000$ \\
\quad G+G & $0.000 \pm 0.000$ & $0.000 \pm 0.000$ & $0.000 \pm 0.000$ & $0.000 \pm 0.000$ & $0.000 \pm 0.000$ & $0.000 \pm 0.000$ \\
\quad He+MS & $0.016 \pm 0.005$ & $0.002 \pm 0.002$ & $0.004 \pm 0.003$ & $0.011 \pm 0.004$ & $0.011 \pm 0.004$ & $0.016 \pm 0.005$ \\
\quad He+G & $0.005 \pm 0.003$ & $0.004 \pm 0.003$ & $0.004 \pm 0.003$ & $0.003 \pm 0.002$ & $0.003 \pm 0.002$ & $0.003 \pm 0.002$ \\
\quad WD+MS & $0.073 \pm 0.011$ & $0.049 \pm 0.010$ & $0.050 \pm 0.010$ & $0.070 \pm 0.010$ & $0.072 \pm 0.011$ & $0.073 \pm 0.011$ \\
\quad WD+G & $0.013 \pm 0.004$ & $0.017 \pm 0.006$ & $0.017 \pm 0.006$ & $0.022 \pm 0.006$ & $0.021 \pm 0.006$ & $0.021 \pm 0.006$ \\
\quad WD+He & $0.002 \pm 0.002$ & $0.000 \pm 0.000$ & $0.002 \pm 0.002$ & $0.006 \pm 0.003$ & $0.006 \pm 0.003$ & $0.008 \pm 0.004$ \\
\quad WD+WD & $0.002 \pm 0.002$ & $0.004 \pm 0.003$ & $0.004 \pm 0.003$ & $0.000 \pm 0.000$ & $0.000 \pm 0.000$ & $0.000 \pm 0.000$ \\
\quad NS+MS & $0.209 \pm 0.018$ & $0.423 \pm 0.030$ & $0.576 \pm 0.035$ & $0.207 \pm 0.018$ & $0.417 \pm 0.026$ & $0.578 \pm 0.031$ \\
\quad NS+CHeB & $0.011 \pm 0.004$ & $0.006 \pm 0.004$ & $0.004 \pm 0.003$ & $0.032 \pm 0.007$ & $0.011 \pm 0.004$ & $0.008 \pm 0.004$ \\
\quad NS+G & $0.082 \pm 0.011$ & $0.047 \pm 0.010$ & $0.000 \pm 0.000$ & $0.073 \pm 0.011$ & $0.047 \pm 0.009$ & $0.008 \pm 0.004$ \\
\quad NS+He & $0.016 \pm 0.005$ & $0.006 \pm 0.004$ & $0.004 \pm 0.003$ & $0.017 \pm 0.005$ & $0.014 \pm 0.005$ & $0.005 \pm 0.003$ \\
\quad NS+WD & $0.013 \pm 0.004$ & $0.009 \pm 0.004$ & $0.002 \pm 0.002$ & $0.008 \pm 0.004$ & $0.005 \pm 0.003$ & $0.003 \pm 0.002$ \\
\quad NS+NS & $0.105 \pm 0.013$ & $0.045 \pm 0.010$ & $0.009 \pm 0.004$ & $0.104 \pm 0.013$ & $0.058 \pm 0.010$ & $0.015 \pm 0.005$ \\
\quad BH+MS & $0.008 \pm 0.004$ & $0.011 \pm 0.005$ & $0.013 \pm 0.005$ & $0.011 \pm 0.004$ & $0.011 \pm 0.004$ & $0.011 \pm 0.004$ \\
\quad BH+CHeB & $0.000 \pm 0.000$ & $0.000 \pm 0.000$ & $0.000 \pm 0.000$ & $0.000 \pm 0.000$ & $0.000 \pm 0.000$ & $0.000 \pm 0.000$ \\
\quad BH+G & $0.000 \pm 0.000$ & $0.000 \pm 0.000$ & $0.000 \pm 0.000$ & $0.002 \pm 0.002$ & $0.002 \pm 0.002$ & $0.002 \pm 0.002$ \\
\quad BH+He & $0.000 \pm 0.000$ & $0.000 \pm 0.000$ & $0.000 \pm 0.000$ & $0.000 \pm 0.000$ & $0.000 \pm 0.000$ & $0.000 \pm 0.000$ \\
\quad BH+WD & $0.000 \pm 0.000$ & $0.000 \pm 0.000$ & $0.000 \pm 0.000$ & $0.000 \pm 0.000$ & $0.000 \pm 0.000$ & $0.000 \pm 0.000$ \\
\quad BH+NS & $0.010 \pm 0.004$ & $0.021 \pm 0.007$ & $0.022 \pm 0.007$ & $0.005 \pm 0.003$ & $0.023 \pm 0.006$ & $0.021 \pm 0.006$ \\
\quad BH+BH & $0.002 \pm 0.002$ & $0.002 \pm 0.002$ & $0.002 \pm 0.002$ & $0.002 \pm 0.002$ & $0.002 \pm 0.002$ & $0.002 \pm 0.002$ \\
\quad Other & $0.008 \pm 0.004$ & $0.000 \pm 0.000$ & $0.000 \pm 0.000$ & $0.009 \pm 0.004$ & $0.010 \pm 0.004$ & $0.010 \pm 0.004$ \\
\bottomrule
\end{tabular}
\caption{ Fractions for merger outcomes in the `isolated binary' simulations, which are based on the systems in which RLOF was triggered in the inner binary system. Data in this table are based on the high-mass tertiary case, i.e,. $m_3=q_2(m_1+m_2)$; see Table~\ref{table:RLOF_q2p} for the low-mass tertiary case. The fractions in this table are given with respect to the systems undergoing RLOF triggered by the primary or secondary star; the fractions of the latter cases corresponding to {\it all} systems were given in Table~\ref{table:fractions_q2}. Here, `He' refers to a helium-burning star. }
\label{table:RLOF_q2}
\end{minipage}
\begin{minipage}{.99\linewidth}
\begin{tabular}{lcccccc}
\toprule
& \multicolumn{6}{c}{Fraction of RLOF systems ($\star 1$ or $\star$ 2)} \\
& \multicolumn{3}{c}{Triple} &\multicolumn{3}{c}{Binary} \\
\midrule
& \multicolumn{3}{c}{$\sk/\kms$} &\multicolumn{3}{c}{$\sk/\kms$} \\
& 0 & 40 & 265 & 0 & 40 & 265 \\
\midrule
RLOF $\rightarrow$ Merger & $0.745 \pm 0.034$ & $0.843 \pm 0.036$ & $0.915 \pm 0.038$ & $0.711 \pm 0.034$ & $0.830 \pm 0.037$ & $0.909 \pm 0.038$ \\
\quad MS+MS & $0.066 \pm 0.010$ & $0.066 \pm 0.010$ & $0.066 \pm 0.010$ & $0.041 \pm 0.008$ & $0.041 \pm 0.008$ & $0.041 \pm 0.008$ \\
\quad CHeB+MS & $0.022 \pm 0.006$ & $0.017 \pm 0.005$ & $0.019 \pm 0.005$ & $0.018 \pm 0.005$ & $0.018 \pm 0.005$ & $0.015 \pm 0.005$ \\
\quad CHeB+G & $0.003 \pm 0.002$ & $0.003 \pm 0.002$ & $0.005 \pm 0.003$ & $0.003 \pm 0.002$ & $0.003 \pm 0.002$ & $0.005 \pm 0.003$ \\
\quad G+MS & $0.054 \pm 0.009$ & $0.061 \pm 0.010$ & $0.058 \pm 0.010$ & $0.055 \pm 0.010$ & $0.057 \pm 0.010$ & $0.055 \pm 0.009$ \\
\quad G+CHeB & $0.000 \pm 0.000$ & $0.000 \pm 0.000$ & $0.000 \pm 0.000$ & $0.000 \pm 0.000$ & $0.000 \pm 0.000$ & $0.000 \pm 0.000$ \\
\quad G+G & $0.000 \pm 0.000$ & $0.000 \pm 0.000$ & $0.000 \pm 0.000$ & $0.000 \pm 0.000$ & $0.000 \pm 0.000$ & $0.000 \pm 0.000$ \\
\quad He+MS & $0.017 \pm 0.005$ & $0.019 \pm 0.005$ & $0.020 \pm 0.006$ & $0.010 \pm 0.004$ & $0.016 \pm 0.005$ & $0.013 \pm 0.005$ \\
\quad He+G & $0.009 \pm 0.004$ & $0.009 \pm 0.004$ & $0.008 \pm 0.004$ & $0.003 \pm 0.002$ & $0.003 \pm 0.002$ & $0.005 \pm 0.003$ \\
\quad WD+MS & $0.068 \pm 0.010$ & $0.067 \pm 0.010$ & $0.072 \pm 0.011$ & $0.072 \pm 0.011$ & $0.081 \pm 0.011$ & $0.072 \pm 0.011$ \\
\quad WD+G & $0.022 \pm 0.006$ & $0.019 \pm 0.005$ & $0.020 \pm 0.006$ & $0.024 \pm 0.006$ & $0.019 \pm 0.006$ & $0.021 \pm 0.006$ \\
\quad WD+He & $0.006 \pm 0.003$ & $0.006 \pm 0.003$ & $0.006 \pm 0.003$ & $0.003 \pm 0.002$ & $0.003 \pm 0.002$ & $0.007 \pm 0.003$ \\
\quad WD+WD & $0.000 \pm 0.000$ & $0.000 \pm 0.000$ & $0.000 \pm 0.000$ & $0.000 \pm 0.000$ & $0.000 \pm 0.000$ & $0.000 \pm 0.000$ \\
\quad NS+MS & $0.217 \pm 0.018$ & $0.411 \pm 0.025$ & $0.570 \pm 0.030$ & $0.212 \pm 0.019$ & $0.409 \pm 0.026$ & $0.596 \pm 0.031$ \\
\quad NS+CHeB & $0.024 \pm 0.006$ & $0.006 \pm 0.003$ & $0.002 \pm 0.002$ & $0.034 \pm 0.007$ & $0.015 \pm 0.005$ & $0.008 \pm 0.004$ \\
\quad NS+G & $0.076 \pm 0.011$ & $0.050 \pm 0.009$ & $0.003 \pm 0.002$ & $0.077 \pm 0.011$ & $0.060 \pm 0.010$ & $0.005 \pm 0.003$ \\
\quad NS+He & $0.016 \pm 0.005$ & $0.009 \pm 0.004$ & $0.006 \pm 0.003$ & $0.018 \pm 0.005$ & $0.006 \pm 0.003$ & $0.003 \pm 0.002$ \\
\quad NS+WD & $0.013 \pm 0.004$ & $0.013 \pm 0.004$ & $0.006 \pm 0.003$ & $0.005 \pm 0.003$ & $0.010 \pm 0.004$ & $0.003 \pm 0.002$ \\
\quad NS+NS & $0.107 \pm 0.013$ & $0.050 \pm 0.009$ & $0.017 \pm 0.005$ & $0.106 \pm 0.013$ & $0.044 \pm 0.008$ & $0.016 \pm 0.005$ \\
\quad BH+MS & $0.011 \pm 0.004$ & $0.011 \pm 0.004$ & $0.011 \pm 0.004$ & $0.011 \pm 0.004$ & $0.011 \pm 0.004$ & $0.011 \pm 0.004$ \\
\quad BH+CHeB & $0.000 \pm 0.000$ & $0.000 \pm 0.000$ & $0.000 \pm 0.000$ & $0.000 \pm 0.000$ & $0.000 \pm 0.000$ & $0.000 \pm 0.000$ \\
\quad BH+G & $0.000 \pm 0.000$ & $0.002 \pm 0.002$ & $0.000 \pm 0.000$ & $0.002 \pm 0.002$ & $0.002 \pm 0.002$ & $0.002 \pm 0.002$ \\
\quad BH+He & $0.002 \pm 0.002$ & $0.000 \pm 0.000$ & $0.000 \pm 0.000$ & $0.000 \pm 0.000$ & $0.000 \pm 0.000$ & $0.000 \pm 0.000$ \\
\quad BH+WD & $0.000 \pm 0.000$ & $0.000 \pm 0.000$ & $0.000 \pm 0.000$ & $0.000 \pm 0.000$ & $0.000 \pm 0.000$ & $0.000 \pm 0.000$ \\
\quad BH+NS & $0.006 \pm 0.003$ & $0.019 \pm 0.005$ & $0.019 \pm 0.005$ & $0.007 \pm 0.003$ & $0.019 \pm 0.006$ & $0.021 \pm 0.006$ \\
\quad BH+BH & $0.002 \pm 0.002$ & $0.000 \pm 0.000$ & $0.002 \pm 0.002$ & $0.000 \pm 0.000$ & $0.002 \pm 0.002$ & $0.002 \pm 0.002$ \\
\quad Other & $0.005 \pm 0.003$ & $0.005 \pm 0.003$ & $0.005 \pm 0.003$ & $0.010 \pm 0.004$ & $0.010 \pm 0.004$ & $0.008 \pm 0.004$ \\
\bottomrule
\end{tabular}
\caption{ Similar to Table~\ref{table:RLOF_q2}, here based on the lower-mass-tertiary simulations.}
\label{table:RLOF_q2p}
\end{minipage}
\end{table*}

As described in detail in \S~\ref{sect:meth:bse}, we continue the evolution of the inner binary after RLOF occurs in the inner binary using the binary population synthesis code \textsc{BSE}. In Tables~\ref{table:RLOF_q2} and \ref{table:RLOF_q2p} for the high- and low-mass tertiary simulations, respectively, we show the fractions for merger outcomes of the subsequent `isolated binary' simulations. These fractions are with respect to the systems undergoing RLOF triggered by the primary or secondary star; the fractions of the latter cases corresponding to {\it all} systems were given in Tables~\ref{table:fractions_q2} and \ref{table:fractions_q2p} for the high- and low-mass tertiary cases, respectively. 

The evolution of the inner binary after RLOF in \textsc{BSE} is driven by several processes, including mass transfer, CE evolution, and SNe. There is a high likelihood, of $\sim 0.7$ to $\sim 0.9$ of the RLOF systems depending on $\sk$, that the inner binary eventually merges. It is evident from Tables~\ref{table:RLOF_q2} and \ref{table:RLOF_q2p} that a large number of merger outcomes are possible. Generally, there are little differences in the fractions between the `triple' and `binary' cases. This can be attributed to the generally only small differences in the properties of the inner binaries when RLOF is triggered (see \S~\ref{sect:results:orbit} below). 

A dominant merger channel is a NS-MS merger, with a fraction of up to $\sim 0.6$ of RLOF systems if $\sk=265\,\kms$. The resulting stars, Thorne-$\mathrm{\dot{Z}}$ytkow objects \citep{1977ApJ...212..832T}, are also found in isolated binary evolution studies \citep[e.g.,][]{1995MNRAS.274..461B}; our results show that this channel is also possible (and relatively likely) in triples. These objects could also evolve to become X-ray binaries, which are also believed to form in triples (without taking into account isolated binary evolution) through the mass-loss induced eccentric LK mechanism \citep{2013ApJ...766...64S,2016ApJ...822L..24N}.

The channel of most interest here is the merger of two NSs, which occurs for a relatively large fraction of $\sim 0.1$ of RLOF systems if $\sk=0\,\kms$, but drops quickly to a tenth of this fraction, to $\sim 0.01$, if $\sk=265\,\kms$. As expected, the kick dispersion has a large impact on the fraction of NS-NS mergers. Also, these fraction are mostly sensitive to $\sk$, and do not depend very strongly on the assumption of the tertiary mass ratio distribution (it should be taken into account, however, that the latter does affect the overall fraction of RLOF systems).

\subsubsection{Mergers from systems with unbound tertiaries}
\label{sect:results:fractions:unbound}

\begin{table}
\begin{minipage}{.99\linewidth}
\begin{tabular}{lcccccc}
\toprule
& \multicolumn{4}{c}{Fraction of unbound triples (stable inner binary)} \\
& \multicolumn{3}{c}{Triple} \\
\midrule
& \multicolumn{3}{c}{$\sk/\kms$} \\
& 0 & 40 & 265 & \\
\midrule
Tertiary unbound \\
$\rightarrow$ inner binary merger & $0.476 \pm 0.041$ & $0.931 \pm 0.055$ & $0.969 \pm 0.057$ &  \\
\quad MS+MS & $0.000 \pm 0.000$ & $0.076 \pm 0.016$ & $0.078 \pm 0.016$ &  \\
\quad CHeB+MS & $0.000 \pm 0.000$ & $0.003 \pm 0.003$ & $0.003 \pm 0.003$ &  \\
\quad CHeB+G & $0.000 \pm 0.000$ & $0.000 \pm 0.000$ & $0.000 \pm 0.000$ &  \\
\quad G+MS & $0.000 \pm 0.000$ & $0.023 \pm 0.009$ & $0.024 \pm 0.009$ &  \\
\quad G+CHeB & $0.000 \pm 0.000$ & $0.000 \pm 0.000$ & $0.000 \pm 0.000$ &  \\
\quad G+G & $0.000 \pm 0.000$ & $0.000 \pm 0.000$ & $0.000 \pm 0.000$ &  \\
\quad He+MS & $0.000 \pm 0.000$ & $0.043 \pm 0.012$ & $0.041 \pm 0.012$ &  \\
\quad He+G & $0.000 \pm 0.000$ & $0.003 \pm 0.003$ & $0.003 \pm 0.003$ &  \\
\quad WD+MS & $0.000 \pm 0.000$ & $0.109 \pm 0.019$ & $0.112 \pm 0.019$ &  \\
\quad WD+G & $0.000 \pm 0.000$ & $0.016 \pm 0.007$ & $0.017 \pm 0.008$ &  \\
\quad WD+He & $0.000 \pm 0.000$ & $0.000 \pm 0.000$ & $0.000 \pm 0.000$ &  \\
\quad WD+WD & $0.000 \pm 0.000$ & $0.000 \pm 0.000$ & $0.000 \pm 0.000$ &  \\
\quad NS+MS & $0.164 \pm 0.024$ & $0.513 \pm 0.041$ & $0.600 \pm 0.045$ &  \\
\quad NS+CHeB & $0.003 \pm 0.003$ & $0.036 \pm 0.011$ & $0.037 \pm 0.011$ &  \\
\quad NS+G & $0.000 \pm 0.000$ & $0.026 \pm 0.009$ & $0.003 \pm 0.003$ &  \\
\quad NS+He & $0.000 \pm 0.000$ & $0.007 \pm 0.005$ & $0.003 \pm 0.003$ &  \\
\quad NS+WD & $0.000 \pm 0.000$ & $0.003 \pm 0.003$ & $0.000 \pm 0.000$ &  \\
\quad NS+NS & $0.252 \pm 0.030$ & $0.013 \pm 0.007$ & $0.007 \pm 0.005$ &  \\
\quad BH+MS & $0.000 \pm 0.000$ & $0.000 \pm 0.000$ & $0.000 \pm 0.000$ &  \\
\quad BH+CHeB & $0.000 \pm 0.000$ & $0.000 \pm 0.000$ & $0.000 \pm 0.000$ &  \\
\quad BH+G & $0.000 \pm 0.000$ & $0.000 \pm 0.000$ & $0.000 \pm 0.000$ &  \\
\quad BH+He & $0.000 \pm 0.000$ & $0.000 \pm 0.000$ & $0.000 \pm 0.000$ &  \\
\quad BH+WD & $0.000 \pm 0.000$ & $0.000 \pm 0.000$ & $0.000 \pm 0.000$ &  \\
\quad BH+NS & $0.056 \pm 0.014$ & $0.053 \pm 0.013$ & $0.034 \pm 0.011$ &  \\
\quad BH+BH & $0.000 \pm 0.000$ & $0.000 \pm 0.000$ & $0.000 \pm 0.000$ &  \\
\quad Other & $0.000 \pm 0.000$ & $0.007 \pm 0.005$ & $0.007 \pm 0.005$ &  \\
\bottomrule
\end{tabular}
\caption{  Fractions for merger outcomes in the `isolated binary' simulations, which are based on the systems in which the tertiary becomes unbound from the inner binary due to a SNe event.  Data in this table are based on the high-mass tertiary case, i.e,. $m_3=q_2(m_1+m_2)$; see Table~\ref{table:unbound_q2p} for the low-mass tertiary case. The fractions in this table are given with respect to the systems in which the tertiary becomes unbound, but with the inner binary still bound; the latter fraction is given in the last row of Table~\ref{table:fractions_q2}. }
\label{table:unbound_q2}
\end{minipage}
\begin{minipage}{.99\linewidth}
\begin{tabular}{lcccccc}
\toprule
& \multicolumn{4}{c}{Fraction of unbound triples (stable inner binary)} \\
& \multicolumn{3}{c}{Triple} \\
\midrule
& \multicolumn{3}{c}{$\sk/\kms$} \\
& 0 & 40 & 265 & \\
\midrule
Tertiary unbound \\
$\rightarrow$ inner binary merger & $0.513 \pm 0.044$ & $0.583 \pm 0.220$ & $0.333 \pm 0.333$ &  \\
\quad MS+MS & $0.000 \pm 0.000$ & $0.000 \pm 0.000$ & $0.000 \pm 0.000$ &  \\
\quad CHeB+MS & $0.000 \pm 0.000$ & $0.000 \pm 0.000$ & $0.000 \pm 0.000$ &  \\
\quad CHeB+G & $0.000 \pm 0.000$ & $0.000 \pm 0.000$ & $0.000 \pm 0.000$ &  \\
\quad G+MS & $0.000 \pm 0.000$ & $0.000 \pm 0.000$ & $0.000 \pm 0.000$ &  \\
\quad G+CHeB & $0.000 \pm 0.000$ & $0.000 \pm 0.000$ & $0.000 \pm 0.000$ &  \\
\quad G+G & $0.000 \pm 0.000$ & $0.000 \pm 0.000$ & $0.000 \pm 0.000$ &  \\
\quad He+MS & $0.000 \pm 0.000$ & $0.000 \pm 0.000$ & $0.000 \pm 0.000$ &  \\
\quad He+G & $0.000 \pm 0.000$ & $0.000 \pm 0.000$ & $0.000 \pm 0.000$ &  \\
\quad WD+MS & $0.000 \pm 0.000$ & $0.000 \pm 0.000$ & $0.000 \pm 0.000$ &  \\
\quad WD+G & $0.000 \pm 0.000$ & $0.000 \pm 0.000$ & $0.000 \pm 0.000$ &  \\
\quad WD+He & $0.000 \pm 0.000$ & $0.000 \pm 0.000$ & $0.000 \pm 0.000$ &  \\
\quad WD+WD & $0.000 \pm 0.000$ & $0.000 \pm 0.000$ & $0.000 \pm 0.000$ &  \\
\quad NS+MS & $0.204 \pm 0.028$ & $0.333 \pm 0.167$ & $0.000 \pm 0.000$ &  \\
\quad NS+CHeB & $0.000 \pm 0.000$ & $0.000 \pm 0.000$ & $0.000 \pm 0.000$ &  \\
\quad NS+G & $0.000 \pm 0.000$ & $0.000 \pm 0.000$ & $0.000 \pm 0.000$ &  \\
\quad NS+He & $0.000 \pm 0.000$ & $0.000 \pm 0.000$ & $0.000 \pm 0.000$ &  \\
\quad NS+WD & $0.000 \pm 0.000$ & $0.000 \pm 0.000$ & $0.000 \pm 0.000$ &  \\
\quad NS+NS & $0.242 \pm 0.030$ & $0.000 \pm 0.000$ & $0.000 \pm 0.000$ &  \\
\quad BH+MS & $0.000 \pm 0.000$ & $0.000 \pm 0.000$ & $0.000 \pm 0.000$ &  \\
\quad BH+CHeB & $0.000 \pm 0.000$ & $0.000 \pm 0.000$ & $0.000 \pm 0.000$ &  \\
\quad BH+G & $0.000 \pm 0.000$ & $0.000 \pm 0.000$ & $0.000 \pm 0.000$ &  \\
\quad BH+He & $0.000 \pm 0.000$ & $0.000 \pm 0.000$ & $0.000 \pm 0.000$ &  \\
\quad BH+WD & $0.000 \pm 0.000$ & $0.000 \pm 0.000$ & $0.000 \pm 0.000$ &  \\
\quad BH+NS & $0.067 \pm 0.016$ & $0.250 \pm 0.144$ & $0.333 \pm 0.333$ &  \\
\quad BH+BH & $0.000 \pm 0.000$ & $0.000 \pm 0.000$ & $0.000 \pm 0.000$ &  \\
\quad Other & $0.000 \pm 0.000$ & $0.000 \pm 0.000$ & $0.000 \pm 0.000$ &  \\
\bottomrule
\end{tabular}
\caption{  Similar to Table~\ref{table:unbound_q2p}, here based on the lower-mass-tertiary simulations. }
\label{table:unbound_q2p}
\end{minipage}
\end{table}

In addition to considering the subsequent `isolated binary' evolution of systems that undergo RLOF, we also consider the `isolated binary' evolution of systems in which the tertiary becomes unbound from the system due to a SNe event, but the inner binary remains bound. In the latter case, the inner binary can subsequently merge due to `isolated binary' evolution, which we take into account by evolving these systems with \textsc{BSE} (see \S~\ref{sect:meth:bse}). Similarly to Tables~\ref{table:RLOF_q2} and \ref{table:RLOF_q2p}, we show in Tables~\ref{table:unbound_q2} and \ref{table:unbound_q2p} the outcome fractions of several merger channels for these `unbound tertiary' systems. Evidently, in this case there are no equivalent `binary' systems (since this channel originates exclusively from triples). The fractions in these tables are given with respect to the systems in which the tertiary becomes unbound, but with a bound inner binary; the latter fractions with respect to {\it all} systems are given in the bottom rows in Tables~\ref{table:fractions_q2} and \ref{table:fractions_q2p} for the high-mass and low-mass tertiaries, respectively.  Note that the latter fraction is very small for the low-mass tertiary simulations and non-zero $\sk$. 

Similarly to RLOF-induced mergers, mergers originating from unbound tertiary systems are dominated by NS-MS mergers, in particular for non-zero $\sk$. The fraction of NS-NS mergers is relatively high at $\sim 0.25$ for $\sk=0\,\kms$, but rapidly decreases with increasing $\sk$, dropping to $\sim 0.01$ for $\sk=265\,\kms$ in the high-mass tertiary case, and to zero (within our statistical certainty) for $\sk=265\,\kms$ in the low-mass tertiary case.

\subsection{Orbital properties}
\label{sect:results:orbit}

\begin{figure}
\center
\includegraphics[scale = 0.48, trim = 5mm 0mm 0mm 0mm]{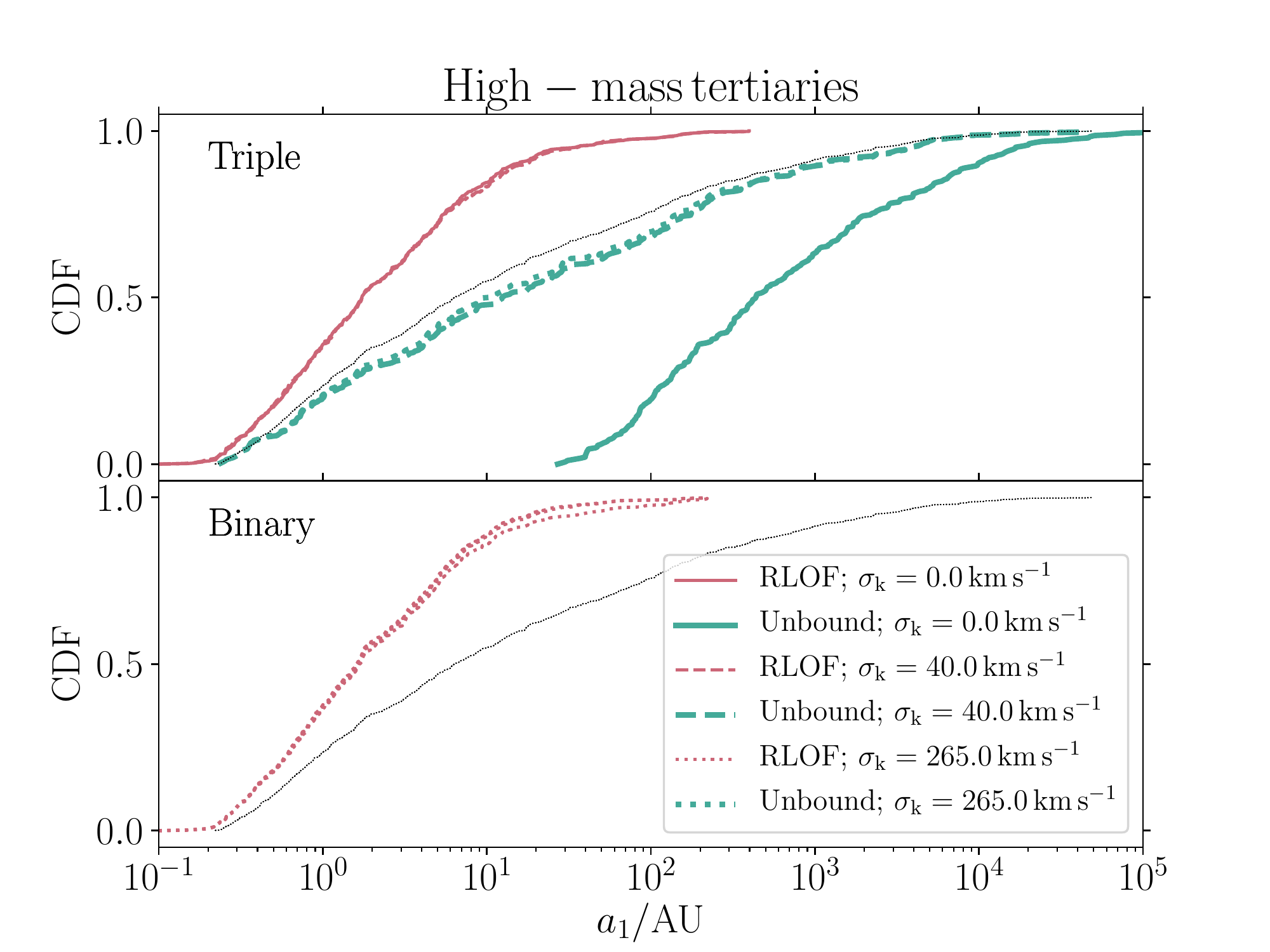}
\includegraphics[scale = 0.48, trim = 5mm 0mm 0mm 0mm]{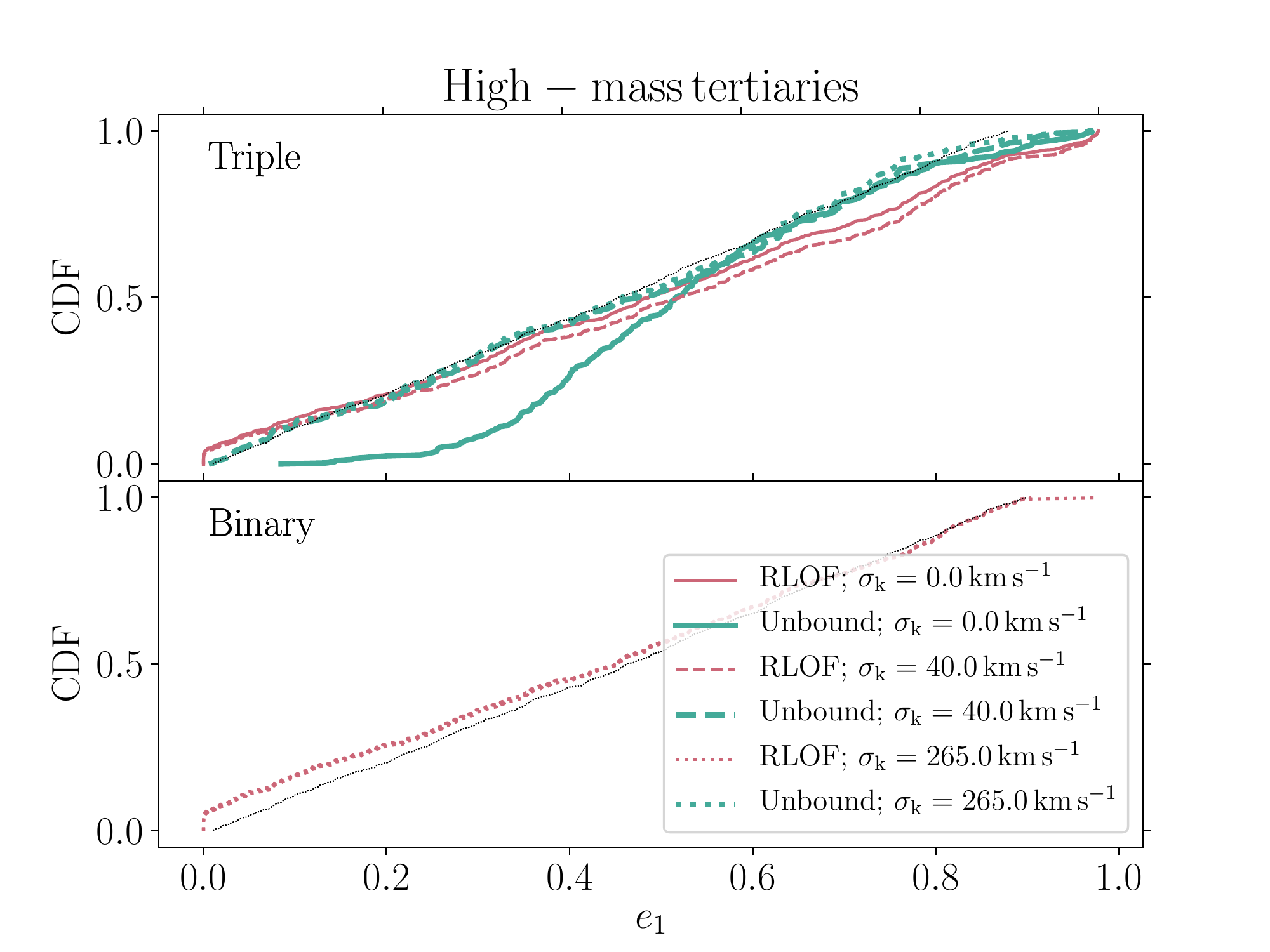}
\caption { Distributions of the inner binary semimajor axis $a_1$ (top panels) and eccentricity $e_1$ (lower panels) at the moment of the onset of RLOF (thin red lines), and the unbinding of the tertiary star (thick green lines). In each set of panels, the top (bottom) panel corresponds to the triple (binary) case. Evidently, the `unbound tertiary' channel does not apply in the binary case. Results here are shown for the high-mass tertiary simulations; refer to \F~\ref{fig:sma_e_q2p} for the low-mass tertiary case. Different line styles correspond to simulations with different $\sk$: lines are solid, dashed and dotted for $\sk=0$, 40, and $265\,\kms$, respectively. The thin dotted black lines show the initial distributions of all systems in the simulations. }
\label{fig:sma_e_q2}
\end{figure}

\begin{figure}
\center
\includegraphics[scale = 0.48, trim = 5mm 0mm 0mm 0mm]{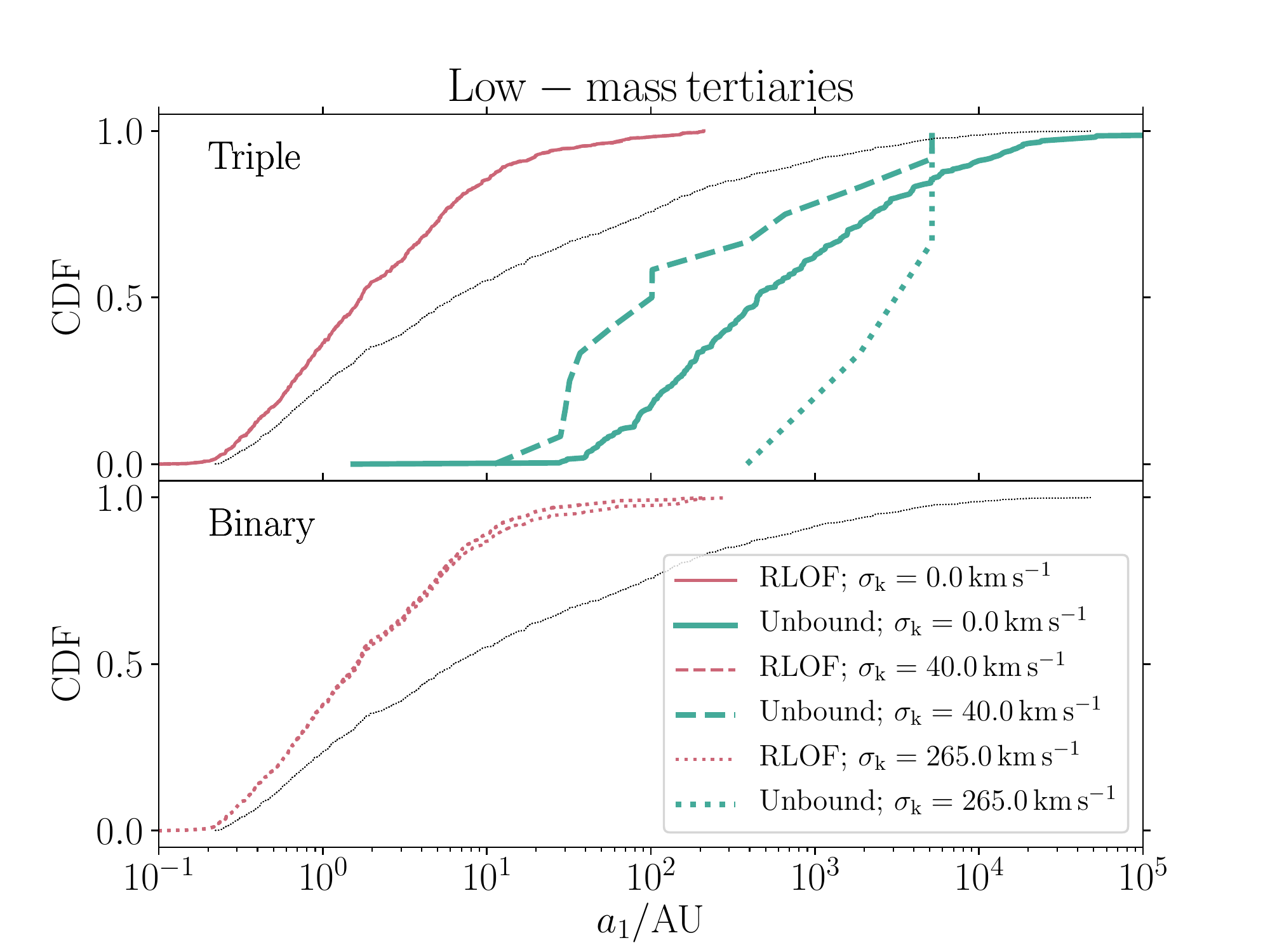}
\includegraphics[scale = 0.48, trim = 5mm 0mm 0mm 0mm]{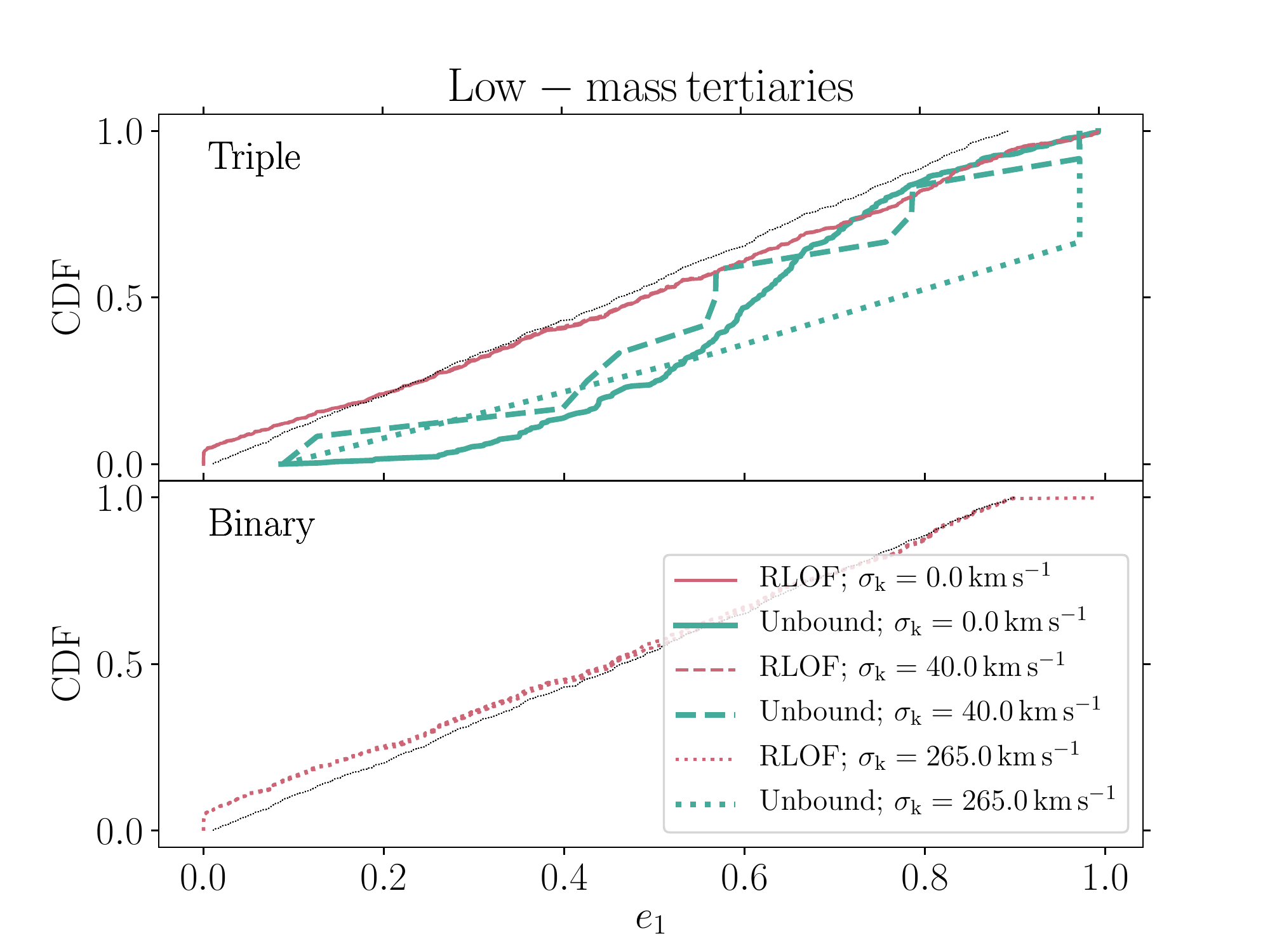}
\caption { Similar to \F~\ref{fig:sma_e_q2}, here for the low-mass tertiary simulations. Note that the number of unbound systems in the low-mass tertiary simulations for $\sk>0\,\kms$ is very small (see the bottom row in Table~\ref{table:fractions_q2p}). }
\label{fig:sma_e_q2p}
\end{figure}

We further discuss the RLOF-induced (\S~\ref{sect:results:fractions:RLOF}) and tertiary unbound (\S~\ref{sect:results:fractions:unbound}) channels by showing in Figs~\ref{fig:sma_e_q2} and \ref{fig:sma_e_q2p} the distributions of the inner binary semimajor axis $a_1$ and eccentricity $e_1$ at the moment of the onset of RLOF, and the unbinding of the tertiary star. Here, Figs~\ref{fig:sma_e_q2} and \ref{fig:sma_e_q2p} correspond to the high- and low-mass tertiary cases, respectively. 

As expected, RLOF-induced systems tend to have significantly smaller inner binary semimajor axes compared to all systems (compare the red and black lines in the figures), with RLOF systems having a median of $a_1\sim 1\,\au$, compared to $a_1\sim 10\,\au$ for all systems overall. Also, SNe kicks have very little impact on the orbital properties RLOF systems, as expected (since in our simulations RLOF typically occurs before stars evolve to compact objects). In addition, for the RLOF systems there are little differences in terms of $a_1$ between the triple and binary cases. There is some difference in terms of the eccentricity -- in the triple case, RLOF-induced systems tend to have slightly higher eccentricities compared to the binary case. This can be explained by eccentricity excitation by LK oscillations. There are no noticeable differences in the orbital properties of RLOF systems between the high- and low-mass tertiary simulations. 

The `tertiary unbound' systems show typically larger semimajor axes compared to all systems (compare the green and black lines in the figures). With higher kicks, the semimajor axes tend to be larger compared to without ($\sk=0\,\kms$); this can be understood by noting that kicks tend to unbind the inner binary if it is wide, so, for the inner binary to remain bound, the inner binary semimajor axis should be smaller. Note that the number of unbound systems in the low-mass tertiary simulations for $\sk>0\,\kms$ are very small (see the bottom row in Table~\ref{table:fractions_q2p}), causing the jagged behavior in \F~\ref{fig:sma_e_q2p}.

\subsection{Delay-time distributions}
\label{sect:results:DTD}

\begin{figure}
\center
\includegraphics[scale = 0.4, trim = 15mm 0mm 0mm 0mm]{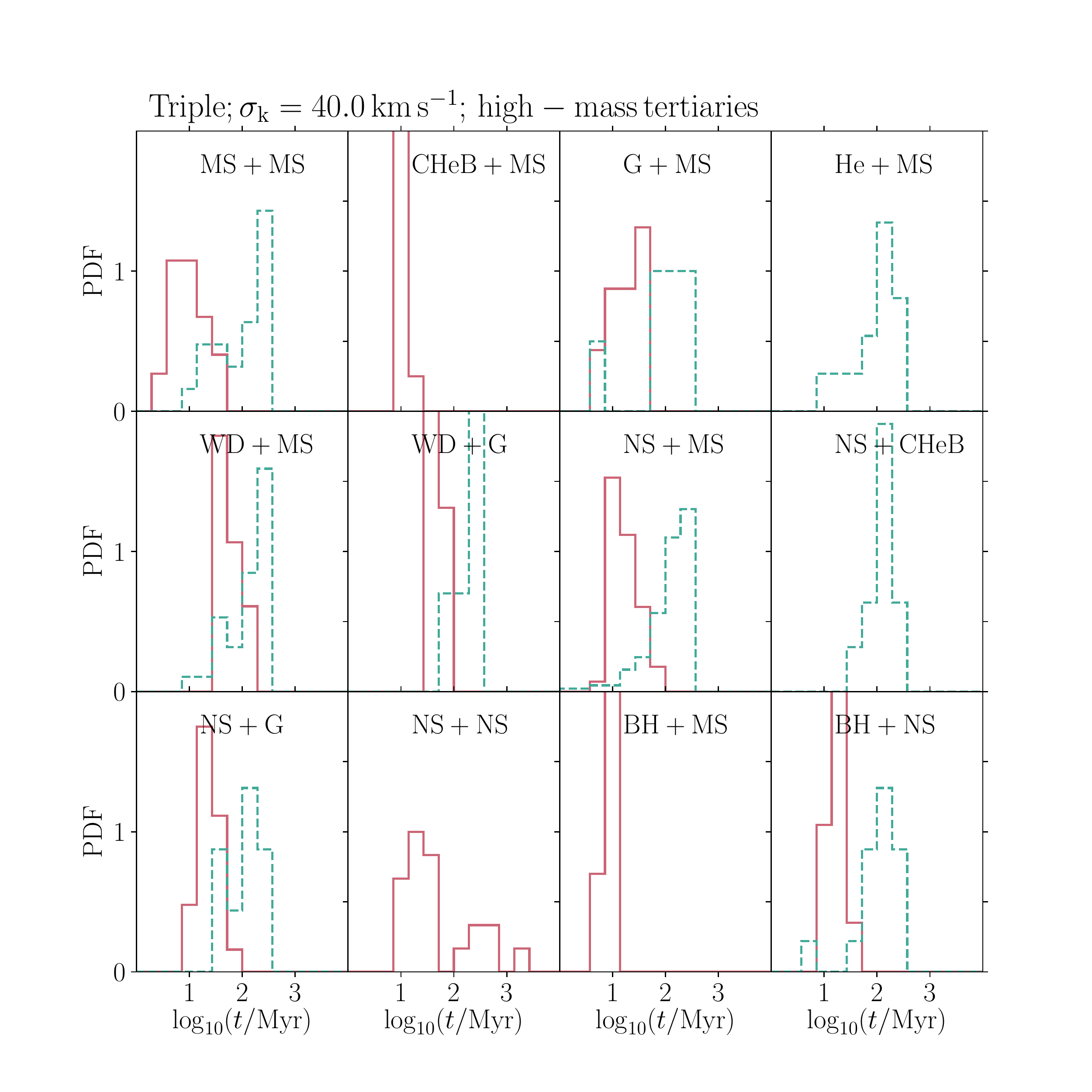}
\caption { Delay-time distributions (DTDs) for various merger channels in the triple simulations. Refer to \F~\ref{fig:DTD_binary} for results from the corresponding binary case. We include the RLOF-induced (red lines), and tertiary unbound (dashed green lines) cases. Each panel corresponds to a different channel, indicated in the top left. Data apply to the simulations with $\sk=40\,\kms$, and the high-mass tertiary assumption.  }
\label{fig:DTD_triple}
\end{figure}

\begin{figure}
\center
\includegraphics[scale = 0.4, trim = 15mm 0mm 0mm 0mm]{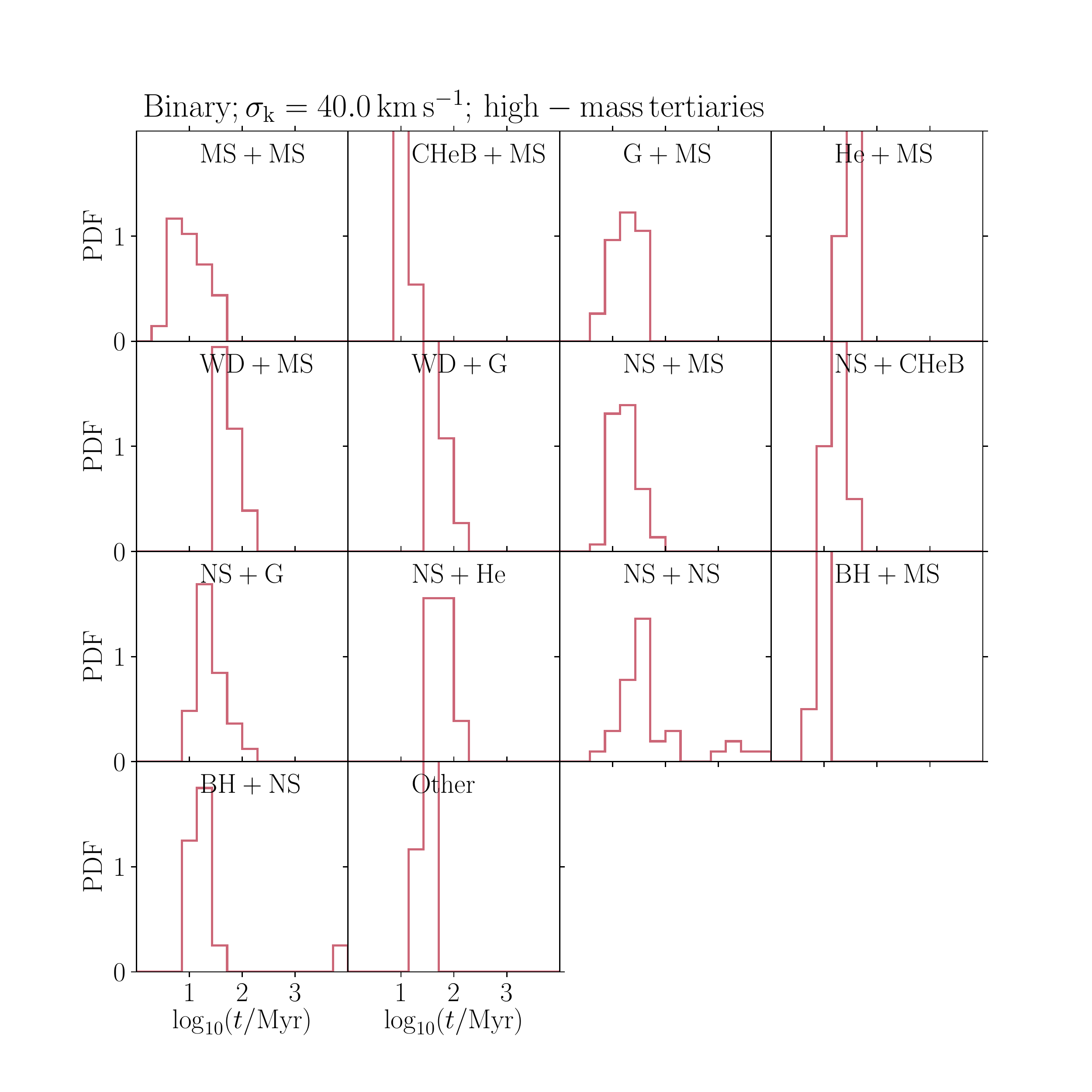}
\caption { Similar to \F~\ref{fig:DTD_triple}, here for the binary case. Note that the tertiary unbound case does not apply here. }
\label{fig:DTD_binary}
\end{figure}

In Figs~\ref{fig:DTD_triple} and \ref{fig:DTD_binary}, we show delay-time distributions (DTDs) for various merger channels for the triple and binary cases, respectively. We include the RLOF-induced (red lines), and tertiary unbound (dashed green lines) cases. Each panel corresponds to a different channel, labeled in the top left, and we show results only for channels with a significant number of systems ($\geq 5$). Data apply to the simulations with $\sk=40\,\kms$, and the high-mass tertiary assumption.  

Generally, RLOF-induced mergers occur earlier than tertiary unbound mergers. This can be attributed to the generally tighter orbits (see \S~\ref{sect:results:orbit}). There are generally no large differences between the triple (\F~\ref{fig:DTD_triple}) and binary (\F~\ref{fig:DTD_binary}) RLOF-induced DTDs, which can be explained by the similarities in the initial orbits. 

Of most interest here are the NS-NS mergers. Most of these mergers occur within $\sim 100\,\mathrm{Myr}$, although there is a small tail with delay times of several Gyr. 

The DTDs for simulations with different parameters (low-mass tertiary, and different $\sk$; not shown here) are qualitatively similar to those in Figs~\ref{fig:DTD_triple} and \ref{fig:DTD_binary}.

\subsection{Merger rates}
\label{sect:results:rates}
Here, we compute the NS-NS merger rates based on the merger fractions from our simulations. To convert merger fractions to absolute rates, we start with the local star formation rate (SFR) density, which we take to be $R_{\mathrm{SFR}} = 0.025\times 10^9\,\msun\,\pgpy$ \citep{2011MNRAS.415.1815B}. This is the mass of all stars formed per unit volume and time (we are agnostic about the type of galaxy in which the stars are formed and consider the local Universe only; this is justified by the typically short delay times, see \S~\ref{sect:results:DTD}). We assume that all stellar systems consist of either single, binary or triple stars (and ignore high-order systems). Consider a population of stars with $\nsys$ stellar systems. The number of single, binary, and triple systems is then $\alhs \nsys$, $\alhb \nsys$, and $\alht \nsys$, respectively, and where $\alhs+\alhb+\alht=1$. We assume $\alhs=0.19$, $\alhb=0.56$, and $\alht=0.25$ \citep{2014ApJS..215...15S}, independent of mass. We remark that other recent observational studies \citep[e.g.,][]{2017ApJS..230...15M} indicate that these fractions could be more heavily biased towards more high-multiplicity systems for higher primary masses. 

We assume a \citet{2001MNRAS.322..231K} initial mass function (IMF), i.e., $\mathrm{d}N/\mathrm{d} m \propto m^{-\alpha}$, where $\alpha=0.3$ for $0.01<m/\msun<0.08$, $\alpha=1.3$ for $0.08<m/\msun<0.5$, and $\alpha=2.3$ for $m>0.5\,\msun$\footnote{The slope of $-2.3$ here is inconsistent with the assumed slope for the massive stars in our simulations (primary masses $8<m_1/\msun<50$; slope $-2.35$). However, the difference in slope is very small. }. A single-star population with $N_{\mathrm{s}}$ stars with such an IMF has a total mass of
\begin{align}
M_{\mathrm{s}} = \int_{m_{\mathrm{low}}}^{m_{\mathrm{up}}} m \frac{\mathrm{d} N}{\mathrm{d} m} \, \mathrm{d} m = N_{\mathrm{s}} \tilde{M},
\end{align}
where $N_{\mathrm{s}}$ is the number of (single) stars, $m_{\mathrm{low}}=0.01\,\msun$ and $m_{\mathrm{up}}=150\,\msun$, and we calculate $\tilde{M}$ to be $\tilde{M} \simeq 0.38\,\msun$ for a \citet{2001MNRAS.322..231K} IMF. In our mixed population with single, binary, and triple stars, the mass contribution from single stars is therefore $\alhs \nsys \tilde{M}$. Assuming a flat mass ratio distribution (i.e., flat in $q_1 \equiv m_2/m_1$), the binaries have a mass contribution which is approximately $\left(1+\frac{1}{2} \right ) \tilde{M} \alhb \nsys = \frac{3}{2}  \alhb \nsys \tilde{M}$. For triples, the mass contribution depends on our assumption on the tertiary mass ratio; we assumed distributions that are flat in either $q_2\equiv m_3/(m_1+m_2)$, or in $q_2'=m_3/m_2$. In the former case, the mass contribution from triples is $\left ( \frac{3}{2} + \frac{1}{2} \frac{3}{2} \right )\tilde{M} \alht \nsys = \frac{9}{4} \alht \tilde{M} \nsys$. In the latter case, it is $\left ( \frac{3}{2} + \frac{1}{2} \frac{1}{2} \right )\tilde{M} \alht \nsys = \frac{7}{4} \alht \tilde{M} \nsys$. Adding up the contributions from all hierarchies, the total mass of our population is (in the high-mass tertiary case)
\begin{align}
M_{\mathrm{tot}} \simeq \left ( \alhs + \frac{3}{2} \alhb + \frac{9}{4} \alht \right ) \nsys \tilde{M} = \left ( 1+ \frac{1}{2} \alhb + \frac{5}{4} \alht \right ) \nsys \tilde{M}.
\end{align}
The number of binaries/triples is therefore given by
\begin{align}
\label{eq:N_bin_tr}
N &= \alpha \nsys = \frac{\alpha M_{\mathrm{tot}}}{  \left ( 1+ \frac{1}{2} \alhb + \frac{5}{4} \alht \right )\tilde{M}},
\end{align}
where $\alpha=\alhb$ for binaries, and $\alpha=\alht$ for triples. The corresponding occurrence rates (per unit volume and time) are obtained by replacing $M_{\mathrm{tot}}$ in equation~(\ref{eq:N_bin_tr}) by $R_{\mathrm{SFR}}$. 

To convert the total occurrence rate of binaries/triples to (NS-NS) merger rates, we need to take into account the fraction of systems that were taken into account in the simulations compared to all astrophysically-occurring systems (which we denote as $f_{\mathrm{calc}}$), and the actual merger fractions in the simulations (which we denote as $f_{\mathrm{merge}}$). Since we did not restrict the orbital separation distributions in the Monte Carlo simulations (see \S~\ref{sect:IC}), the calculated fraction $f_{\mathrm{calc}}$ is determined solely by the mass cutoffs. Specifically, we restricted the primary mass to the range $8<m_1/\msun<50$ and the secondary mass to $4<m_2/\msun<50$, whereas there was no restriction on $m_3$ (within the mass ranges of the IMF of \citealt{2001MNRAS.322..231K}). For the adopted \citet{2001MNRAS.322..231K} IMF, this implies that $f_{\mathrm{calc}} \simeq 2.6 \times 10^{-3}$, which applies to both triples and binaries. The merger fractions $f_{\mathrm{merge}}$ can be inferred from Tables~\ref{table:fractions_q2}, \ref{table:fractions_q2p}, \ref{table:RLOF_q2}, \ref{table:RLOF_q2p}, \ref{table:unbound_q2}, and \ref{table:unbound_q2p}.

\begin{figure}
\center
\includegraphics[scale = 0.48, trim = 5mm 0mm 0mm 0mm]{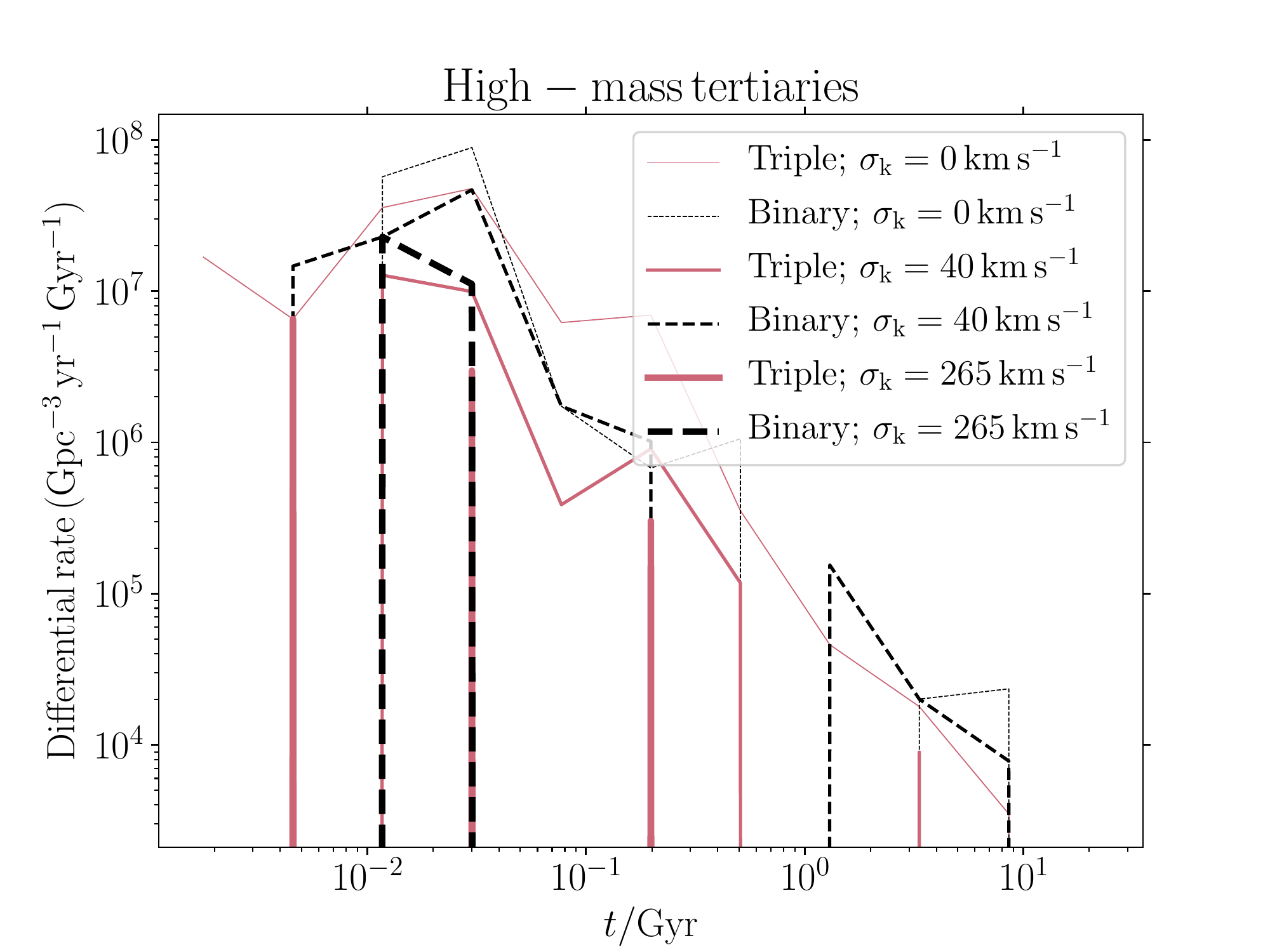}
\caption { DTD distributions for NS-NS mergers, normalized to absolute merger rates (i.e., the integrated curves correspond to the rates given in Table~\ref{table:rates}). Red solid (black dashed) lines correspond to triples (binaries). Line thickness increases with increasing $\sigma_\kick$. }
\label{fig:DTD_rates}
\end{figure}

\begin{figure}
\center
\includegraphics[scale = 0.48, trim = 5mm 0mm 0mm 0mm]{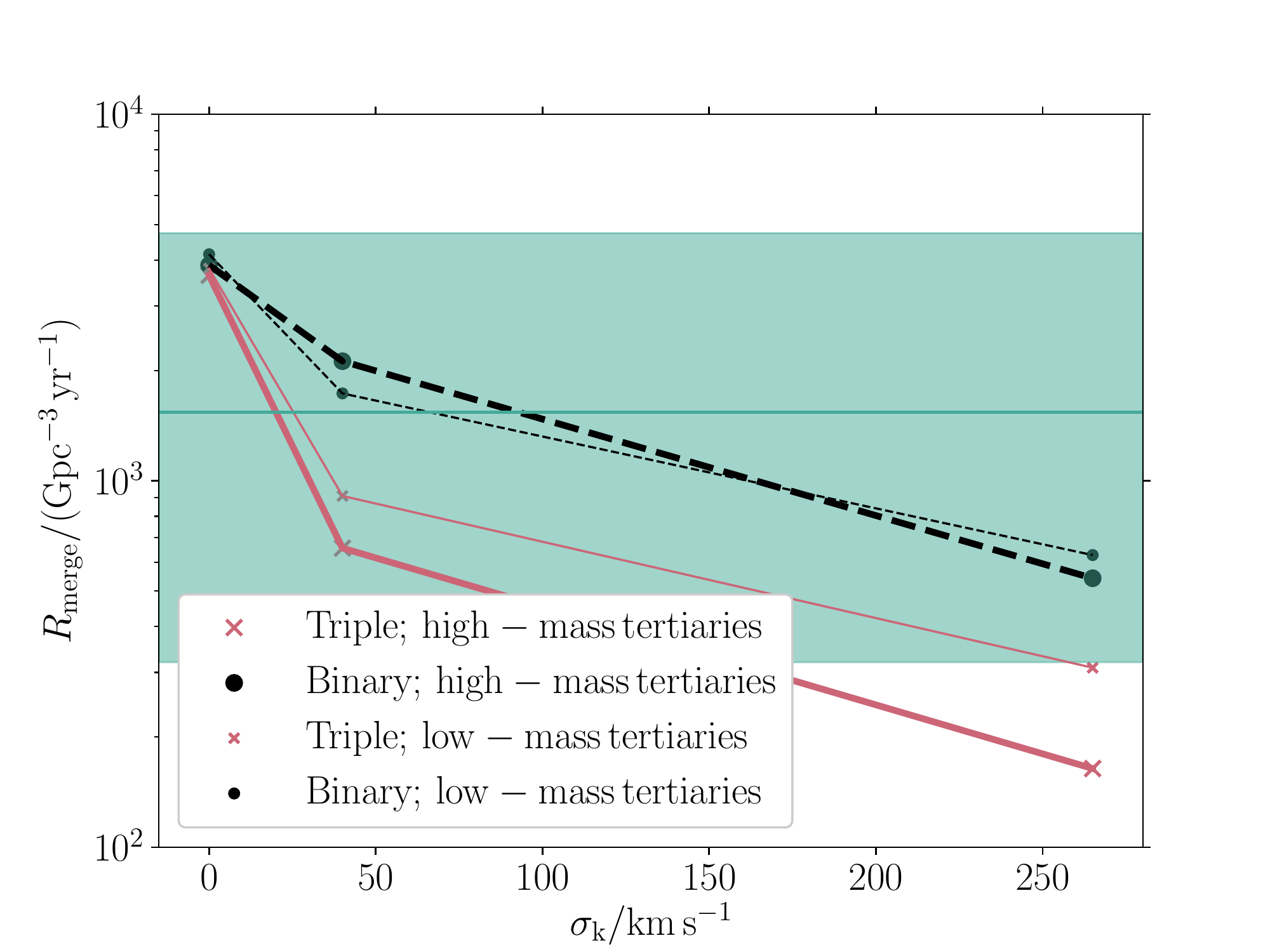}
\caption { Total NS-NS rates plotted as a function of $\sigma_\kick$ (the same data are quoted in Table~\ref{table:rates}). Red solid (black dashed) lines correspond to triples (binaries), and the thick (thin) lines correspond to the high (low)-mass tertiary cases. The horizontal solid green line (with green error regions) shows the LIGO rate inferred from GW170817 \cite{2017PhRvL.119p1101A}. }
\label{fig:rate_summary}
\end{figure}

The merger rate for binaries/triples (number per unit volume and time) is then given by
\begin{align}
\label{eq:rates}
R_{\mathrm{merge}} \simeq \frac{\alpha R_{\mathrm{SFR}}}{ \left ( 1+ \frac{1}{2} \alhb + \frac{5}{4} \alht \right ) \tilde{M}} f_{\mathrm{calc}} f_{\mathrm{merge}}.
\end{align}
Note that this equation applies to the high-mass tertiary case; in the low-mass tertiary case, the factor $\frac{5}{4}$ in equation~(\ref{eq:rates}) should be replaced by $\frac{3}{4}$. The resulting NS-NS merger rates for all our simulations are given in Table~\ref{table:rates}. We also give the BH-NS merger rates in the same table. We do not include BH-BH rates, since the associated fractions $f_{\mathrm{merge}}$ in the simulations are at the noise level. The NS-NS DTD distributions normalized to the rates are shown in \F~\ref{fig:DTD_rates}, and the total rates are plotted as a function of $\sigma_\kick$ in \F~\ref{fig:rate_summary}. We discuss our rates further in \S~\ref{sect:discussion:rates}.

\begin{table}
\begin{tabular}{lcccccc}
\toprule
& \multicolumn{6}{c}{Rate ($\pgpy$)} \\
& \multicolumn{3}{c}{Triple} &\multicolumn{3}{c}{Binary} \\
\midrule
& \multicolumn{3}{c}{$\sk/\kms$} &\multicolumn{3}{c}{$\sk/\kms$} \\
& 0 & 40 & 265 & 0 & 40 & 265 \\
\midrule
NS-NS \\
\midrule
High-mass tertiary & \\
\quad RLOF		& 1741 & 551 & 109 & 3874 & 2119 & 542 \\
\quad Unbound		& 1893 & 104 & 54 & --- & --- & --- \\
\quad Total		& 3634 & 655 & 164 & 3874 & 2119 & 542 \\
\midrule
Low-mass tertiary & \\
\quad RLOF		& 1937 & 909 & 309 & 4149 & 1731 & 627 \\
\quad Unbound		& 1856 & 0 & 0 & --- & --- & --- \\
\quad Total		& 3793 & 909 & 309 & 4149 & 1731 & 627 \\
\midrule
\midrule
BH-NS \\
\midrule
High-mass tertiary & \\
\quad RLOF		& 166 & 257 & 268 & 186 & 840 & 759 \\
\quad Unbound		& 421 & 423 & 263 & --- & --- & --- \\
\quad Total		& 586 & 680 & 531 & 186 & 840 & 759 \\
\midrule
Low-mass tertiary & \\
\quad RLOF		& 109 & 346 & 345 & 274 & 747 & 823 \\
\quad Unbound		& 514 & 21 & 0 & --- & --- & --- \\
\quad Total		& 622 & 367 & 345 & 274 & 747 & 823 \\
\midrule
\midrule
NS-MS \\
\midrule
High-mass tertiary & \\
\quad RLOF		& 3464 & 5178 & 7006 & 7710 & 15238 & 20883 \\
\quad Unbound		& 1232 & 4097 & 4650 & --- & --- & --- \\
\quad Total		& 4697 & 9275 & 11656 & 7710 & 15238 & 20883 \\
\midrule
Low-mass tertiary & \\
\quad RLOF		& 3928 & 7475 & 10351 & 8299 & 16088 & 23368 \\
\quad Unbound		& 1564 & 28 & 0 & --- & --- & --- \\
\quad Total		& 5493 & 7504 & 10351 & 8299 & 16088 & 23368 \\
\bottomrule
\end{tabular}
\caption{ Compact object merger rates according to our simulations. Top part: NS-NS mergers; middle part: BH-NS mergers. We also include the rate of NS-MS mergers (Thorne-$\mathrm{\dot{Z}}$ytkow objects) in the bottom part. We include mergers from triples and binaries, and for different kick dispersions $\sk$. The assumed channels are mergers following RLOF, and after the tertiary becomes unbound from the inner binary (only applies to triples). The `Total' row gives the sum of the two channels (note: quoted numbers have been rounded to integers). Results are shown for sets of simulations with the tertiary mass sampled according to $m_3=q_2(m_1+m_2)$ (`high-mass tertiary' case), and according to $m_3=q_2' m_2$ (`low-mass tertiary' case), where both $q_2$ and $q_2'$ have flat distributions. }
\label{table:rates}
\end{table}

\section{Discussion}
\label{sect:discussion}

\subsection{The decoupling assumption}
\label{sect:discussion:decouple}

\begin{figure}
\center
\includegraphics[scale = 0.48, trim = 5mm 0mm 0mm 0mm]{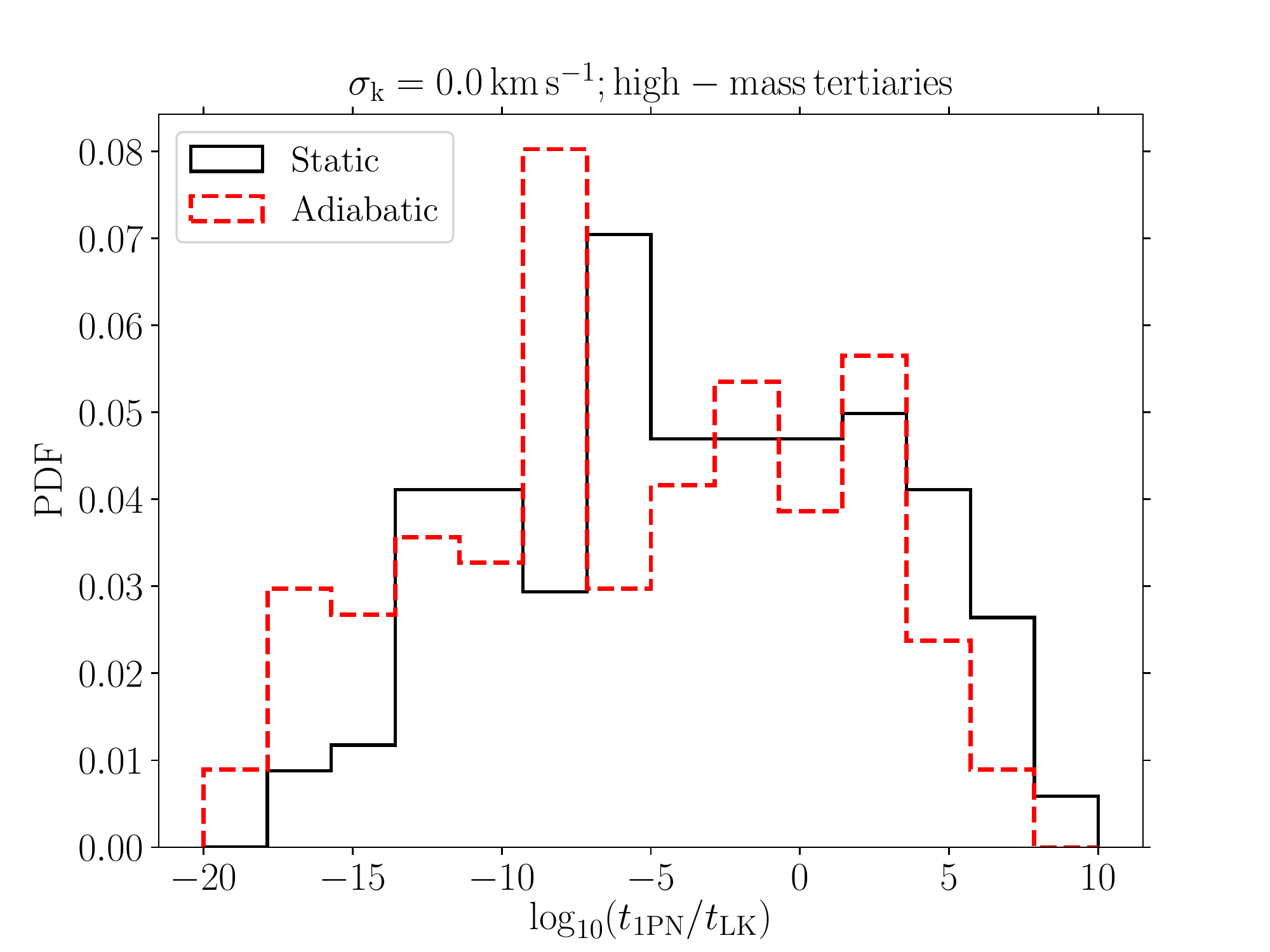}
\includegraphics[scale = 0.48, trim = 5mm 0mm 0mm 0mm]{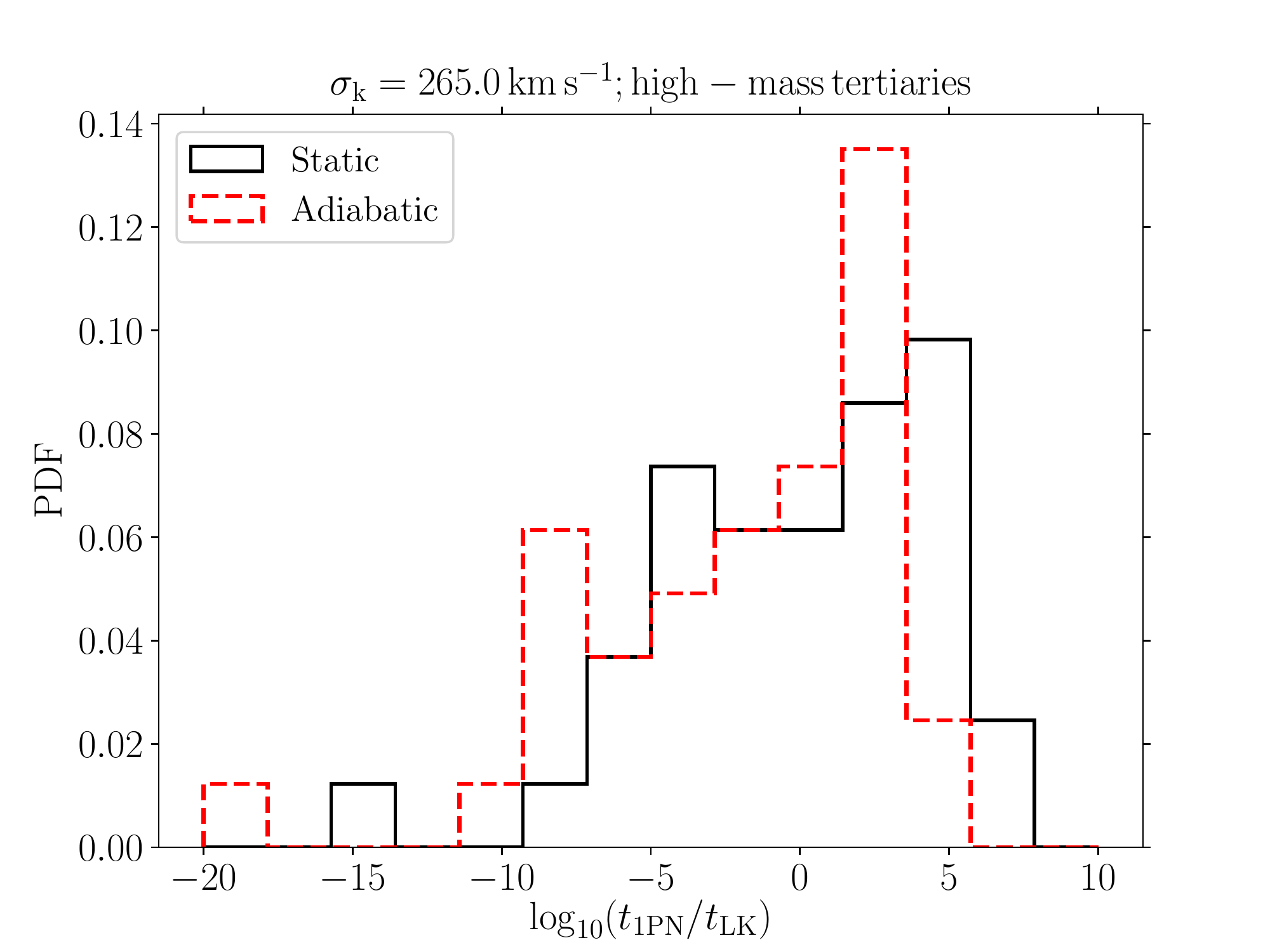}
\caption { Distributions of the ratio $t_{\mathrm{1PN}}/t_{\mathrm{LK}}$ for systems in which RLOF was triggered in the inner binary, but which do not merge within 10 Gyr according to \textsc{BSE}. The top (bottom) panels assume $\sk=0\,\kms$ ($\sk=265\,\kms$), and data are taken from the high-mass tertiary simulations (the distributions are similar for the low-mass tertiary simulations). Two assumptions are made on the outer orbit semimajor axis $a_2$ after 10 Gyr: we either take it to be the value at the onset of RLOF (`static'; black solid lines), or the value if all mass in the inner binary between the time of the onset of RLOF and 10 Gyr were lost adiabatically, i.e., with $a_2(m_1+m_2+m_3)$ constant (`adiabatic'; red dashed lines). }
\label{fig:decouple}
\end{figure}

As mentioned in \S~\ref{sect:meth:bse}, a (strong) assumption that we made in our modeling is that, after RLOF is triggered in the inner binary system, the subsequent evolution of the inner binary is completely decoupled from the tertiary star. This approach was taken for simplicity, and current lack of the available framework to model this phase self-consistently. Here, we briefly investigate this assumption, by considering the systems that, after RLOF is triggered in the inner binary, survive the binary evolution, i.e., do not merge within 10 Gyr. 

To evaluate whether or not the inner system can be decoupled from the tertiary, we compute for these systems the ratio of the timescale $t_{\mathrm{1PN}}$ for the lowest-order post-Newtonian (PN) precession \citep[e.g.,][]{1972gcpa.book.....W}, to the LK timescale $t_{\mathrm{LK}}$ \citep[e.g.,][]{1999CeMDA..75..125K,2015MNRAS.452.3610A}. If $t_{\mathrm{1PN}}\ll t_{\mathrm{LK}}$, we expect 1PN precession to dominate and the decoupling assumption to be justified, whereas if $t_{\mathrm{1PN}}\gg t_{\mathrm{LK}}$, this is no longer the case \citep[e.g.,][]{2002ApJ...578..775B,2003ApJ...589..605W,2007ApJ...669.1298F,2011ApJ...741...82T,2013ApJ...773..187N,2015MNRAS.447..747L}.

In our simulations, we do not model the subsequent evolution of the tertiary star and orbit after the onset of RLOF. Here, we assume, for simplicity, that the tertiary mass does not change between the onset of RLOF and 10 Gyr, and make two assumptions on the outer orbit semimajor axis $a_2$ after 10 Gyr: we either take it to be the value at the onset of RLOF, or the value if all mass in the inner binary between the time of the onset of RLOF and 10 Gyr were lost adiabatically, i.e., with $a_2(m_1+m_2+m_3)$ constant. 

In \F~\ref{fig:decouple}, we show the resulting distributions of the ratio $t_{\mathrm{1PN}}/t_{\mathrm{LK}}$, for two values of $\sk$, and the high-mass tertiary simulations. As shown, there is a non-negligible fraction of systems with $t_{\mathrm{1PN}}/t_{\mathrm{LK}}\gg1$, in which case the decoupling assumption is not well justified. The fraction of these systems depends on $\sk$; the distribution of $t_{\mathrm{1PN}}/t_{\mathrm{LK}}$ shifts to larger values with increasing $\sk$. 

This indicates that not all systems are well decoupled from the tertiary. This implies there is a systematic error in our results, in particular of the NS-NS merger rates. We also note that we here only considered the surviving systems; the systems undergoing RLOF may also not be truly decoupled. At this point, it is difficult to estimate whether fully self-consistent modeling would lead to lower or higher merger rates. Such an endeavor is left for future work.

\subsection{Rates}
\label{sect:discussion:rates}
Here, we comment on our NS-NS merger rates (see \S~\ref{sect:results:rates}). First, as also discussed above, it should be noted that there are likely significant systematic errors in our rates as a consequence of simplifications made in the modeling. Therefore, our results should be interpreted as estimates.

Interestingly, and most importantly, the triple rates, although always lower, are typically comparable to the binary rates (see Table~\ref{table:rates}). Both the triple and binary rates are highly sensitive to the SNe kick dispersion $\sk$, as expected. However, the triple rates are even more sensitive to $\sk$ than the binary rates. This can be explained by the fact that a third star adds more possibilities for the system to become unbound, and this effect increases in importance with increasing $\sk$. It should also be mentioned that we assumed a conservative triple fraction ($\alht=0.25$) compared to the binary fraction ($\alhb=0.56$). If a higher triple fraction were assumed relative to the binary fraction, for which there is observational evidence for massive multiple systems \citep{2017ApJS..230...15M}, then the importance of triple systems to the NS-NS merger rate is even larger. In particular, the ratio of the triple to the binary merger rate is $\propto \alht/\alhb$ (see equation~\ref{eq:rates}); therefore, if we assumed, e.g., $\alht=0.5$ and $\alhb=0.4$ instead of $\alht=0.25$ and $\alhb=0.56$, this ratio would increase by a factor of $\simeq 3$. 

Our rates vary from typically several hundred to several thousand $\pgpy$, strongly depending on $\sk$. The rate inferred by GW170817 is $1540_{-1220}^{+3200}\,\pgpy$ \citep{2017PhRvL.119p1101A}, which falls well within our ranges (for both triples and binaries). Based on observations of short gamma-ray bursts, \citet{2012MNRAS.425.2668C} find a rate of $\sim 8-1800\,\pgpy$, \citet{2013ApJ...767..140P} of $\sim500-1500\,\pgpy$, \citet{2014MNRAS.437..649S} of $\sim 92-1154\,\pgpy$, and \citet{2015ApJ...815..102F} of $\sim90-1850\,\pgpy$, also consistent with our rates. 

Theoretical studies based on isolated binary evolution typically find NS-NS merger rates of several tens to hundreds $\pgpy$, up to order thousand in extreme cases. For example, \citet{2016ApJ...819..108B} found rates of $\sim 50-150\,\pgpy$, and \citet{2018MNRAS.481.1908K} found rates of $\sim10-400\,\pgpy$ (the upper limit being `rather optimistic'). The rates are highly sensitive to the CE $\alpha$ parameter, as shown by \citet{2018MNRAS.480.2011G}, who found rates up to several hundred $\pgpy$ and up to $\sim 10^3\,\pgpy$ if a high CE $\alpha$ value is assumed, and \citet{2018MNRAS.474.2937C}, who found similar results. Another crucial ingredient is the assumption on natal kicks \citep[e.g.,][]{2019MNRAS.482.2234G}.

Generally, isolated binary evolution studies find rates that are lower than our `binary' rates, as listed in Table~\ref{table:rates}. This can be understood by noting that our `binary' population consists of the inner binaries of a triple population. The requirement of dynamical stability implies that the inner binary separation distribution becomes skewed towards smaller values. As can be seen in \F~\ref{fig:sma_e_q2} (see the black dotted lines), the inner binary semimajor axis distribution of the binaries in our simulations has a median value of $\approx 10\,\au$. In contrast, for a distribution flat in $\log_{10}(a_1)$ without the restriction of dynamical stability imposed by the tertiary, the median semimajor axis is $a_{\mathrm{low}} (a_{\mathrm{up}}/a_{\mathrm{low}})^{1/2} \simeq 32\,\au$ (setting $a_{\mathrm{low}}=0.02\,\au$, and $a_{\mathrm{up}} = 5\times10^4\,\au$, see \F~\ref{fig:sma_e_q2}). A more compact distribution of semimajor axes implies a higher formation rate of double NS binaries, explaining the higher base binary NS-NS merger rate in our simulations. Another way of understanding this trend is by noting that a more compact initial semimajor axis distribution is somewhat analogous to assuming a higher CE efficiency. The latter indeed leads to higher merger rates, of several hundred and up to $\sim 10^3$ $\pgpy$ \citep{2018MNRAS.480.2011G}.

\section{Conclusions}
\label{sect:conclusions}
In this paper, we estimated the rates of mergers of double neutron stars (NSs) in triple systems. We took into account secular, stellar, tidal and binary evolution, and the effects of supernovae (SNe) on the orbits, starting with main-sequence (MS) stars, until the merger of two NSs. We made different assumptions on the properties of the massive triple progenitors, including different kick distributions, and different assumptions on the tertiary mass distribution. Our main conclusions are given below.

\medskip \noindent 1. Contrary to previous studies of the secular and stellar evolution of triples which focused on wide inner binaries that do not interact in the absence of a tertiary star \citep[e.g.,][]{2013MNRAS.430.2262H,2017ApJ...841...77A,2019arXiv190412881H}, we found that the tertiary does not significantly affect the probability of Roche-lobe overflow (RLOF) in the inner binary system. This can be attributed to the fact that we did not restrict to wide, non-interacting systems, but instead considered the entire range of orbital separations. Consequently, the inner binary interacts also in the absence of the tertiary star. Typically, about $60\%$ of systems undergo RLOF in our simulations, for both triple and binary cases (in the latter case, the same inner binary parameters are adopted, but without taking into account the effects of the tertiary star). We modeled the subsequent evolution of RLOF systems using a binary population synthesis code, neglecting the effect of the tertiary. Subsequently, the stars in the inner binary can merge as two NSs after periods of mass transfer and/or common-envelope evolution. Future work should not make the simplifying decoupling assumption, but model the system self-consistently (see \S~\ref{sect:discussion:decouple}). 

\medskip \noindent 2. For $\sim 10$ up to $\sim 50\%$ of systems, the inner and/or outer binary becomes unbound due to the effects of SNe (instantaneous mass loss and/or the effects of SNe kicks). Kicks are typically more important for triples, since a third star adds more possibilities for the system to become unbound, and this effect increases in importance with increasing kick dispersion. In up to $\sim 30\%$ of simulated triples, the inner binary remains bound after a SNe event whereas the tertiary star becomes unbound. The inner binaries of these systems can still potentially merge at a later time due to `isolated binary' evolution. We continued the evolution of these isolated binaries using a binary population synthesis code (and considered these systems to be part of the original `triple' population). The fraction of triple systems without strong interactions is small, typically a few per cent. This can be attributed to the large RLOF fraction (due to the inclusion of tighter systems), and the importance of SNe kicks, especially for NSs.

\medskip \noindent 3. We considered two pathways for NS-NS mergers in triples: following binary interactions after RLOF (possibly induced by the tertiary star through Lidov-Kozai oscillations), or following binary interactions after the inner binary became unbound from the tertiary due to a SNe event. For the equivalent binaries (with the same properties as the inner binaries of the triples that we modeled), we considered only the channel of mergers following binary interactions after RLOF. Our rates (see \F\,\ref{fig:rate_summary} for a visual summary) vary from typically several hundred to several thousand $\pgpy$, and are within the rate estimates of LIGO based on GW170817, $1540_{-1220}^{+3200}\,\pgpy$ \citep{2017PhRvL.119p1101A}. We find that the rates decrease strongly with increasing SNe kick speed, $\sigma_\kick$. Also, the ratio of the triple to binary NS-NS merger rate decreases with increasing $\sigma_\kick$. Our `binary' rates are higher compared to dedicated isolated binary evolution studies. This can be understood from the more compact inner binaries in our simulations, which are the result of the requirement of dynamical stability of the corresponding triple system. 

\medskip \noindent 4. Most of the NS-NS mergers in our models occur relatively early, with a delay-time distribution (DTD) peaked around several tens of Myr. Some mergers can occur at late times, of several Gyr. 

\medskip \noindent 5. We also considered mergers of other types of stars (see Figs. \ref{fig:DTD_triple} and \ref{fig:DTD_binary}). In particular, we found a large fraction of NS-MS mergers in both triple and binary cases. Such mergers result in Thorne-$\mathrm{\dot{Z}}$ytkow objects \citep{1977ApJ...212..832T}, and are also found in dedicated isolated binary evolution studies \citep[e.g.,][]{1995MNRAS.274..461B}. Our results show that this channel is also possible (and relatively likely) in triples; in fact, the fractions of Thorne-$\mathrm{\dot{Z}}$ytkow objects in our simulations are very similar in the triple and binary cases, giving formation rates on the order of several thousand $\pgpy$ (see Table~\ref{table:rates}). Taking into account the occurrence rates of triples and binaries, the relative formation rate of these objects formed in our simulations ranges between $R_\mathrm{T\dot{Z},triple}/R_\mathrm{T\dot{Z},binary} \sim 0.4$ to $\sim 0.6$ (depending on the assumed kick speed and tertiary mass ratio distributions). In addition, we found BH-NS mergers at a rate of typically several hundred $\pgpy$ (see Table~\ref{table:rates}).

\section*{Acknowledgements}
We thank the anonymous referee for a helpful report. A.S.H. gratefully acknowledges support from the Institute for Advanced Study, and the Martin A. and Helen Chooljian Membership. T.A.T. is supported in part by a Simons Foundation Fellowship, an IBM Einstein Fellowship from the Institute for Advanced Study, NSF grant 1313252, and Scialog Scholar grant 24216 from the Research Corporation.

\bibliographystyle{yahapj}
\bibliography{literature}


\end{document}